\documentclass[%
 aip,
 amsmath,amssymb,
 reprint,%
]{revtex4-1}

\usepackage{graphicx}
\usepackage{dcolumn}
\usepackage{bm}

\usepackage[utf8]{inputenc}
\usepackage[T1]{fontenc}
\usepackage{mathptmx}
\usepackage{etoolbox}

\makeatletter
\def\@email#1#2{%
 \endgroup
 \patchcmd{\titleblock@produce}
  {\frontmatter@RRAPformat}
  {\frontmatter@RRAPformat{\produce@RRAP{*#1\href{mailto:#2}{#2}}}\frontmatter@RRAPformat}
  {}{}
}%
\makeatother
\begin{document}

\preprint{AIP/123-QED}

\title{Wavefunction Embedding for Molecular Polaritons}


\author{Fabijan Pavo\v{s}evi\'{c}}
\email{fpavosevic@gmail.com}
\affiliation{Center for Computational Quantum Physics, Flatiron Institute, 162 5th Ave., New York, 10010  NY,  USA}

\author{Angel Rubio}
\email{angel.rubio@mpsd.mpg.de}
\affiliation{Center for Computational Quantum Physics, Flatiron Institute, 162 5th Ave., New York, 10010  NY,  USA}
\affiliation{Max Planck Institute for the Structure and Dynamics of Matter and
Center for Free-Electron Laser Science \& Department of Physics,
Luruper Chaussee 149, 22761 Hamburg, Germany}
\affiliation{Nano-Bio Spectroscopy Group and European Theoretical Spectroscopy Facility (ETSF), Universidad del Pa\'is Vasco (UPV/EHU), Av. Tolosa 72, 20018 San Sebastian, Spain}


\begin{abstract}
Polaritonic chemistry relies on the strong light-matter interaction phenomena for altering the chemical reaction rates inside optical cavities. To explain and to understand these processes, the development of reliable theoretical models is essential. While computationally efficient quantum electrodynamics self-consistent field (QED-SCF) methods, such as quantum electrodynamics density functional theory (QEDFT) needs accurate functionals, quantum electrodynamics coupled cluster (QED-CC) methods provide a systematic increase in accuracy but at much greater cost. To overcome this computational bottleneck, herein we introduce and develop the QED-CC-in-QED-SCF projection-based embedding method that inherits all the favorable properties from the two worlds, computational efficiency and accuracy. 
The performance of the embedding method is assessed by studying some prototypical but relevant reactions, such as methyl transfer reaction, proton transfer reaction, as well as protonation reaction in a complex environment. The results obtained with the new embedding method are in excellent agreement with more expensive QED-CC results. The analysis performed on these reactions indicate that the strong light-matter interaction is very local in nature and that only a small region should be treated at the QED-CC level for capturing important effects due to cavity. This work sets the stage for future developments of polaritonic quantum chemistry methods and it will serve as a guideline for development of other polaritonic embedding models.
\end{abstract}

\maketitle

\section{Introduction}

The ability to control the rates of chemical reactions is the "Holy Grail" sought by chemists, and the strong light-matter interactions created by quantum fluctuations or external pumping in optical or nanoplasmonic cavities offer a non-intrusive way to modulate the rates of chemical reactions. In a newly emerging field of polaritonic chemistry, such strong light-matter coupling effects are utilized to catalyze,~\cite{lather2019cavity,campos2019resonant,climent2019plasmonic} inhibit,~\cite{thomas2016ground,hutchison2012modifying} or modify the overall reaction path~\cite{fregoni2018manipulating,thomas2019tilting} of a given chemical reaction. Naturally, these experimental advances have been accompanied by various theoretical developments~\cite{fregoni2018manipulating,ruggenthaler2018quantum,lacombe2019exact,campos2019resonant,li2021collective,schafer2021shining,sidler2020chemistry,sidler2021perspective} that attempt to provide an insight into fundamental understanding of the strong light-matter interaction as well as for guiding design principles of processes inside cavities.~\cite{ruggenthaler2018quantum}

Like in the case of the conventional electronic structure methods, there are two main approaches to solve the Schr\"odinger equation that accounts for the strong light-matter interaction. One is with the density functional theory, and another is based on the wave function theory. In the quantum electrodynamics density functional theory (QEDFT) approach~\cite{ruggenthaler2014quantum,flick2015kohn,flick2017,schafer2020relevance} both photons and electrons are treated quantum mechanically on the equal footing via a generalized matter-photon exchange-correlation functional. This method is favored for its ability to balance the accuracy and the computational cost, making it suitable for treatment of large molecular systems, and it effectively captures most important effects arising from the strong light-matter coupling.~\cite{flick2018abinito} However, the practical implementation of the QEDFT method so far has similar issues that are inherent to the conventional electronic DFT method, such as problems associated with self-interaction error~\cite{cohen2008insights} and dispersion interactions.~\cite{hermann2017} Despite promising directions in designing exchange-correlation functionals, only relatively small number of the exchange-correlation functionals~\cite{pellegrini2015,Schafere2110464118} for polaritonic problems are currently available.
An alternative to QEDFT, but in principle more costly, are systematically improvable wave function based methods such as quantum electrodynamics coupled cluster (QED-CC) method.~\cite{mordovina2020polaritonic,haugland2020coupled} 
Although the practical implementations of this method rely on numerous approximations and assumptions with plenty of room for further developments,~\cite{haugland2020coupled} the QED-CC method retains many favorable properties of the electronic CC method such as the size-extensivity,~\cite{bartlett2007coupled} and high accuracy as demonstrated for different chemical processes in optical cavities.~\cite{haugland2021intermolecular,deprince2021cavity,Pavosevic2021,liebenthal2021equationofmotion,pavosevic2021cavity} However, due to a steep polynomial scaling, its applicability is limited to very small molecular systems. 

One way for extending the range of applicability of the QED-CC method can be achieved within the quantum many body embedding approach. In this approach,~\cite{jones2020embedding} only a small region or chemically active subsystem is treated at a high level theory, whereas the rest of the system (environment subsystem) is described with a lower level theory. Among various embedding approaches,~\cite{sun2016quantum,reinhard2019density,jones2020embedding} a particularly popular and robust is the projection-based embedding method.~\cite{manby2012simple,lee2019projection} In this method, the orthogonality of occupied orbitals between the active subsystem and the environment subsystem is achieved via the level shift projection operator, that shifts the occupied orbital energies of the environment subsystem to higher energies.~\cite{manby2012simple} This ensures that the sum of energies of both fragments is equal to the energy of the full system if both fragments are treated at the same level of theory. Therefore, this method is also referred as exact SCF-in-SCF embedding method,~\cite{manby2012simple} where SCF (Self-Consistent Field) can either be the Hartree-Fock (HF) or DFT method. Yet another appealing feature of the projection based embedding is that the correlation energy of the active subsystem is obtained seamlessly without any modification of the post-SCF code. The computational savings comes from the fact that the correlation energy is calculated with a fewer occupied orbitals. For instance, the coupled cluster with singles and doubles (CCSD) scales as $\mathcal{O}(o^2v^4)$, where $o$ and $v$ are the number of occupied and virtual (unoccupied) orbitals, respectively. The projection-based embedding method reduces this scaling
to $\mathcal{O}(v^4)$.~\cite{lee2019projection} Additional computational savings are achieved by selecting the unoccupied (virtual) orbitals that are relevant for the embedded subsystem.~\cite{bennie2015accelerating,bensberg2019automatic,claudino2019simple} In principle, such truncation of the virtual subspace ensures that the computational cost of the embedding region is independent of the system size.~\cite{bennie2015accelerating,claudino2019simple} 

Encouraged by an impressive performance of the projection-based embedding as already implemented in widely used Molpro quantum chemistry software,~\cite{werner2020molpro} as well as by robustness of the QED-CC method, in this work we develop the exact QED-SCF-in-QED-SCF projection-based embedding method as well as the QED-CC-in-QED-SCF method for polaritonic systems. The accuracy and computational performance of the newly developed QED-CC-in-QED-SCF method is verified and benchmarked on the Menshutkin reaction, intramolecular proton transfer reaction in the Z-3-amino-propenal (aminopropenal) molecule, and on the proton binding energy of methanol in the explicit water solvent. The developments and analysis presented in this work highlight the capabilities, versatility, and numerical efficiency of the QED-CC-in-SCF method for accurate description of strong-light matter interaction effect created in optical and nanoplasmonic cavities. It also showcases that only a small part of the system should be treated at a high level of theory for achieving an accurate description of the effects due to cavity. Moreover, this work lays the foundation for developments of other theoretical models in polaritonic chemistry, that will lead to further fundamental understanding of the role of the complex environment in an optical cavity, and it will serve as a guideline and a benchmark for development of other embedding models. To the best of our knowledge, this is the first multi-scale embedding method for treatment of molecular polaritons and it has a tremendous potential for applications to large molecular polaritonic ensembles.

\section{Theory}

The quantum mechanical non-relativistic treatment of interaction between molecules and photons inside an optical cavity can be described by the Pauli-Fierz Hamiltonian.\cite{ruggenthaler2014quantum,ruggenthaler2018quantum} This Hamiltonian (in atomic units unless otherwise stated), within the dipole approximation (we assume the wavelength of the cavity is much larger than the molecule), in the length gauge,~\cite{craig1998,schafer2020relevance} in the coherent state basis,~\cite{ruggenthaler2018quantum,haugland2020coupled} and for a single cavity photon mode (extension to many cavity modes is, in principle, straightforward) reads as
\begin{equation}
\begin{aligned}
    \label{eqn:PF_Hamiltonian_CSB}
    \hat{H}&=h^p_q a_p^q + \frac{1}{2}g^{pq}_{rs} a_{pq}^{rs}+\omega_{\text{cav}}b^{\dagger}b\\\newline
    &-\sqrt{\frac{\omega_\text{cav}}{2}}(\boldsymbol{\lambda} \cdot \Delta\boldsymbol{d})(b^{\dagger}+b)+\frac{1}{2}(\boldsymbol{\lambda} \cdot \Delta\boldsymbol{d})^2.
\end{aligned}
\end{equation}
The first two terms constitute the electronic Hamiltonian (within the Born-Oppenheimer approximation, although non-adiabatic effects can also be incorporated~\cite{Hammes-Schiffer19_338,Hammes-Schiffer20_4222,pavosevic2021multicomponent}) that is defined in terms of the second-quantized electronic excitation operator $a_{p_1p_2...p_n}^{q_1q_2...q_n}=a_{q_1}^{\dagger}a_{q_2}^{\dagger}...a_{q_n}^{\dagger}a_{p_n}...a_{p_2}a_{p_1}$ where $a^{\dagger}/a$ are fermionic creation/annihilation operators. Moreover, $h^p_q=\langle q|\hat{h}^\text{e}|p\rangle$ and $g^{pq}_{rs}=\langle rs|pq\rangle$ is the core electronic Hamiltonian matrix element and the two-electron repulsion tensor element, respectively. The indices $i,j,k,l,...$, $a,b,c,d,...$, and $p,q,r,s,...$ denote occupied, unoccupied, and general electronic spin orbitals, respectively. The third term denotes the oscillation of the single cavity mode with a fundamental frequency $\omega_\text{cav}$ expressed in terms of bosonic creation/annihilation ($b^{\dagger}$/$b$) operators. The fourth term describes the dipolar coupling between electronic and photonic degrees of freedom, where $\boldsymbol{\lambda}$ is the light-matter coupling strength vector, and $\Delta\boldsymbol{d}=\boldsymbol{d}-\langle\boldsymbol{d}\rangle$ is the dipole fluctuation operator that denotes the change of the molecular dipole operator with respect to its expectation value. $\boldsymbol{\lambda}$ defines the strength of the light-matter coupling which in turn depends on the dielectric constant $\epsilon_0$ and the effective quantization volume ($V_\text{eff}$) as $\lambda = 1/\sqrt{\epsilon_0 V_\text{eff}}$.~\cite{climent2019plasmonic,Galego2019} This $V_\text{eff}$ depends on the specific experimental realizations of the optical cavity, and experiments in picocavity setups with effective volume $<1$ nm$^3$ (which corresponds to values of $\lambda>0.1$ a.u) have been recently achieved.~\cite{benz20216, Richard_Lacroix_2020, Liu2021,griffiths2021} Lastly, the fifth term in Eq.~\eqref{eqn:PF_Hamiltonian_CSB} describes the dipole self energy.~\cite{rokaj2018,schafer2020relevance,Taylor2020} The role of the dipole self energy term is to ensure the origin invariance of the Hamiltonian and its boundness from below (i.e., for avoiding the "ground-state catastrophe").~\cite{rokaj2018,schafer2020relevance,andolina2020theory}

\begin{figure*}[ht]
  \centering
  \includegraphics[width=6.5in]{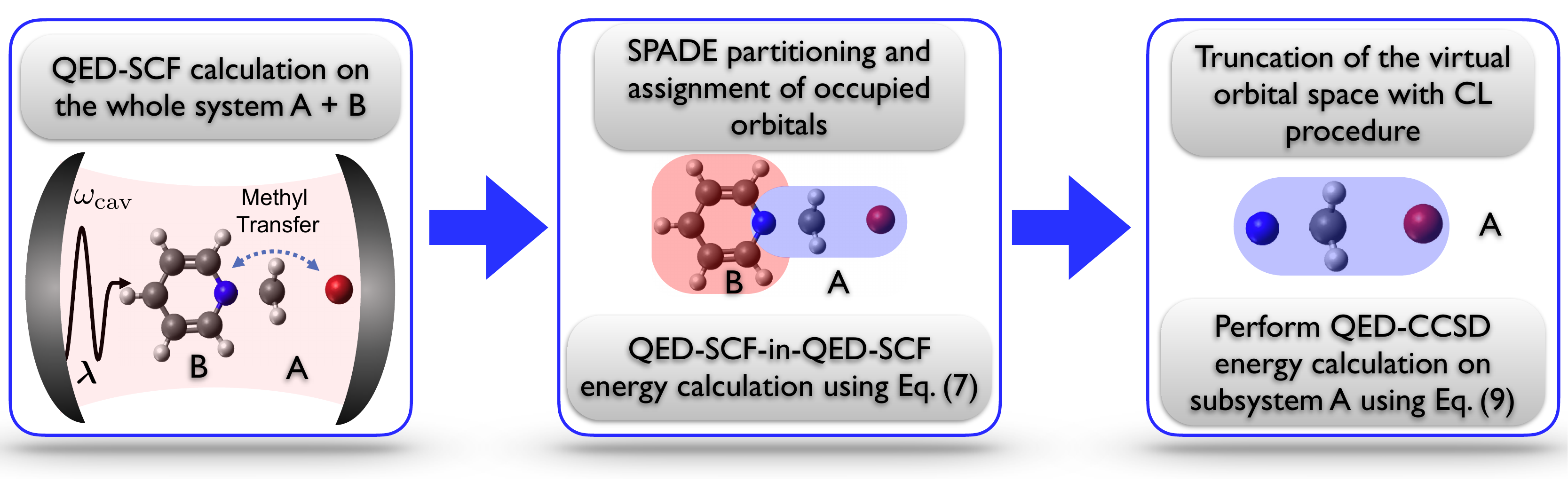}
  \caption{Schematic depiction of the steps performed in the embedding QED-CCSD-in-QED-SCF method.}
  \label{fig:schematic_representation}
\end{figure*}

Just like in conventional electronic structure methods, the usual starting point for the accurate correlated methods is the quantum electrodynamics Hartree-Fock (QED-HF) method,~\cite{haugland2020coupled} in which the wave function ansatz 
\begin{equation}
    \label{eqn:QED-HF}
    |0^{\text{e}}0^{\text{ph}}\rangle=|0^{\text{e}}\rangle\otimes|0^{\text{ph}}\rangle
\end{equation}
is expressed as a direct product between an electronic Slater determinant $|0^{\text{e}}\rangle$ and a photon-number state $|0^{\text{ph}}\rangle$, where the superscripts $\text{e}$ and $\text{ph}$ denote electrons and photons, respectively. Because the QED-HF method treats the electrons and photons as uncorrelated particles that interact through the mean-field potential, the predictions obtained from this method are often inaccurate and unreliable.~\cite{haugland2021intermolecular,pavosevic2021cavity} Within the QED-HF procedure, the correlation effects between electrons can be simply included by adding the exchange-correlation potential to the working QED-HF equations, which defines the QED-DFT method. Note that unlike the QEDFT method, such QED-DFT method will lack the important electron-photon correlation effects. In QED-CC,~\cite{haugland2020coupled} the correlation effects between quantum electrons and photons are accounted through the exponentiated form of the cluster operator
\begin{equation}
    \label{eqn:QED-T}
    \hat{T}=\sum_{\mu,n}t_{\mu,n}a^{\mu}(b^{\dagger})^n
\end{equation}
as
\begin{equation}
    \label{eqn:QED-CC}
    |\Psi_{\text{QED-CC}}\rangle=e^{\hat{T}}|0^{\text{e}}0^{\text{ph}}\rangle.
\end{equation}
The unknown wave function parameters $t_{\mu,n}$ (amplitudes) are determined by solving a set of nonlinear equations
\begin{equation}
    \label{eqn:QED-T-equations}
    \langle0^{\text{e}}0^{\text{ph}}|a_{\mu}(b)^ne^{-\hat{T}}\hat{H}e^{\hat{T}}|0^{\text{e}}0^{\text{ph}}\rangle=\sigma_{\mu,n}.
\end{equation}
Here, $a^\mu=a_\mu^{\dagger}=\{a_{i}^{a},a_{ij}^{ab},...\}$ is the electronic excitation operator, the index $\mu$ is the electronic excitation manifold (rank), and $n$ corresponds to the number of photons. 
Truncation of the cluster operator to include up to single and double electronic excitations along with their interactions with up to two quantum photons is expressed as
\begin{equation}
\begin{aligned}
    \label{eqn:QED-T-22}
    \hat{T}&=t^{i,0}_a a_i^a+t^{0,1}b^\dagger+\frac{1}{4}t^{ij,0}_{ab} a_{ij}^{ab}+t^{i,1}_aa_i^ab^\dagger+\frac{1}{4}t^{ij,1}_{ab} a_{ij}^{ab}b^\dagger\\&+t^{0,2}b^\dagger b^\dagger+t^{i,2}_aa_i^ab^\dagger b^\dagger+\frac{1}{4}t^{ij,2}_{ab} a_{ij}^{ab}b^\dagger b^\dagger,
\end{aligned}
\end{equation}
and defines the QED-CCSD-22 method~\cite{Pavosevic2021,pavosevic2021cavity} that we will simply refer as QED-CCSD in the remainder of this article.

Next, we will briefly describe the QED projection-based quantum embedding technique which works in analogous way to its conventional electronic counterpart. This approach starts by first performing the QED-SCF calculation (QED-HF or QED-DFT) that provides molecular orbitals. These orbitals along with their corresponding density matrices $\boldsymbol{\gamma}$ are then partition into the active subsystem A and the environment subsystem B. Then the energy expression of the subsystem A embedded in subsystem B is
\begin{equation}
\begin{aligned}
    \label{eqn:QED-SCF-in-QED-SCF}
    &E_{\text{QED-SCF-in-QED-SCF}}[\boldsymbol{\gamma}_{\text{emb}}^{\text{A}};\boldsymbol{\gamma}^{\text{A}},\boldsymbol{\gamma}^{\text{B}}]=\\\newline &E_{\text{QED-SCF}}[\boldsymbol{\gamma}_{\text{emb}}^{\text{A}}]+E_{\text{QED-SCF}}[\boldsymbol{\gamma}^{\text{A}}+\boldsymbol{\gamma}^{\text{B}}]-E_{\text{QED-SCF}}[\boldsymbol{\gamma}^{\text{A}}]\\\newline &+\text{tr}[(\boldsymbol{\gamma}_{\text{emb}}^{\text{A}}-\boldsymbol{\gamma}^{\text{A}})\boldsymbol{v}_{\text{emb}}[\boldsymbol{\gamma}^{\text{A}},\boldsymbol{\gamma}^{\text{B}}]]+\mu\text{tr}[\boldsymbol{\gamma}_{\text{emb}}^{\text{A}}\mathbf{P}].
\end{aligned}
\end{equation}
where $\boldsymbol{\gamma}_{\text{emb}}^{\text{A}}$ is the density matrix of the embedded subsystem A, $E_{\text{QED-SCF}}$ is the QED-HF or QED-DFT energy evaluated with a given density, $\mathbf{P}$ is a projector that ensures orthogonality of orbitals between subsystems, and finally $\boldsymbol{v}_{\text{emb}}$ is the embedding potential defined as
\begin{equation}
\begin{aligned}
    \label{eqn:embed-pot}
    \boldsymbol{v}_{\text{emb}}[\boldsymbol{\gamma}^{\text{A}},\boldsymbol{\gamma}^{\text{B}}]=\boldsymbol{\tilde{g}}[\boldsymbol{\gamma}^{\text{A}}+\boldsymbol{\gamma}^{\text{B}}]-\boldsymbol{\tilde{g}}[\boldsymbol{\gamma}^{\text{A}}].
\end{aligned}
\end{equation}
In this equation, $\boldsymbol{\tilde{g}}$ includes all two-electron terms such as the Coulomb, exchange, dipole self energy, as well as the electronic exchange-correlation contributions in case of the QED-DFT method. We note that the sum of energies of both fragments is equal to the energy of the full system if both fragments are treated at the same QED-SCF level of theory.
The described QED-SCF-in-QED-SCF embedding method given by Eq. \eqref{eqn:QED-SCF-in-QED-SCF} sets the stage for the QED-CCSD-in-QED-SCF embedding method, where the active subsystem A is treated using the QED-CCSD method from Eq. \eqref{eqn:QED-T-22} and environment subsystem B is described with the QED-SCF method. Then the QED-CCSD-in-QED-SCF energy is simply obtained by substituting the QED-SCF energy of the subsystem A with the QED-CCSD energy as
\begin{equation}
\begin{aligned}
    \label{eqn:QED-CC-in-QED-SCF}
    &E_{\text{QED-CCSD-in-QED-SCF}}[\Psi_{\text{QED-CCSD}}^{\text{A}};\boldsymbol{\gamma}^{\text{A}},\boldsymbol{\gamma}^{\text{B}}]=\\\newline &E_{\text{QED-CCSD}}[\Psi_{\text{QED-CCSD}}^{\text{A}}]+E_{\text{QED-SCF}}[\boldsymbol{\gamma}^{\text{A}}+\boldsymbol{\gamma}^{\text{B}}]\\\newline&-E_{\text{QED-SCF}}[\boldsymbol{\gamma}^{\text{A}}] +\text{tr}[(\boldsymbol{\gamma}_{\text{emb}}^{\text{A}}-\boldsymbol{\gamma}^{\text{A}})\boldsymbol{v}_{\text{emb}}[\boldsymbol{\gamma}^{\text{A}},\boldsymbol{\gamma}^{\text{B}}]]\\\newline
    &+\mu\text{tr}[\boldsymbol{\gamma}_{\text{emb}}^{\text{A}}\mathbf{P}].
\end{aligned}
\end{equation}
Here, $E_{\text{QED-CCSD}}[\Psi_{\text{QED-CCSD}}^{\text{A}}]$ and $|\Psi_{\text{QED-CCSD}}^{\text{A}}\rangle$ are the QED-CCSD energy and the wave function of the subsystem A. In the remainder of this article we will refer to QED-CCSD-in-QED-SCF simply as QED-CCSD-in-SCF. Schematic depiction of the workflow for the QED-CC-in-QED-SCF method is given in Fig.~\ref{fig:schematic_representation}.

\section{Results}

The QED-CCSD-in-SCF embedding method have been implemented in an in-house version of the Psi4NumPy quantum chemistry software,~\cite{smith2018psi4numpy,githubpsi4numpy} which will be made publicly available in near future. The calculations to study the Menshutkin reaction were performed on the geometries optimized at the conventional electronic MP2/6-31G(d)~\cite{ditchfield1971self,hariharan1973influence,rassolov20016} level using the Orca quantum chemistry software.~\cite{neese2020orca} The calculations to study the proton transfer in aminopropenal molecule were performed on the geometries optimized at the conventional electronic CCSD/cc-pVDZ level.~\cite{dunning1989gaussian} The geometries of the optimized structures (reactants, transition states, and products) are provided in the supplementary material. The cavity effect on the proton binding energy of methanol in an explicit water solvent was performed on geometries obtained from Ref.~\citenum{parravicini2021embedded}.
All of the QED-SCF calculations are carried out with the HF method, as well as with the PBE,~\cite{perdew1996generalized} and hybrid PBE0~\cite{adamo1999toward} and B3LYP~\cite{Parr88_785,Becke88_3098} functionals. The partitioning of the full system into the subsystems A and B is performed with the Subsystem Projected Atomic-orbital Decomposition (SPADE)~\cite{claudino2019automatic} procedure. Truncation of the virtual (unoccupied) orbital space is carried out by employing the Concentric Localization (CL) of orbitals~\cite{claudino2019simple} procedure, which can be viewed as an extension of the SPADE procedure for partitioning of the unoccupied orbitals.

\begin{figure*}[ht!]
  \centering
  \includegraphics[width=6.5in]{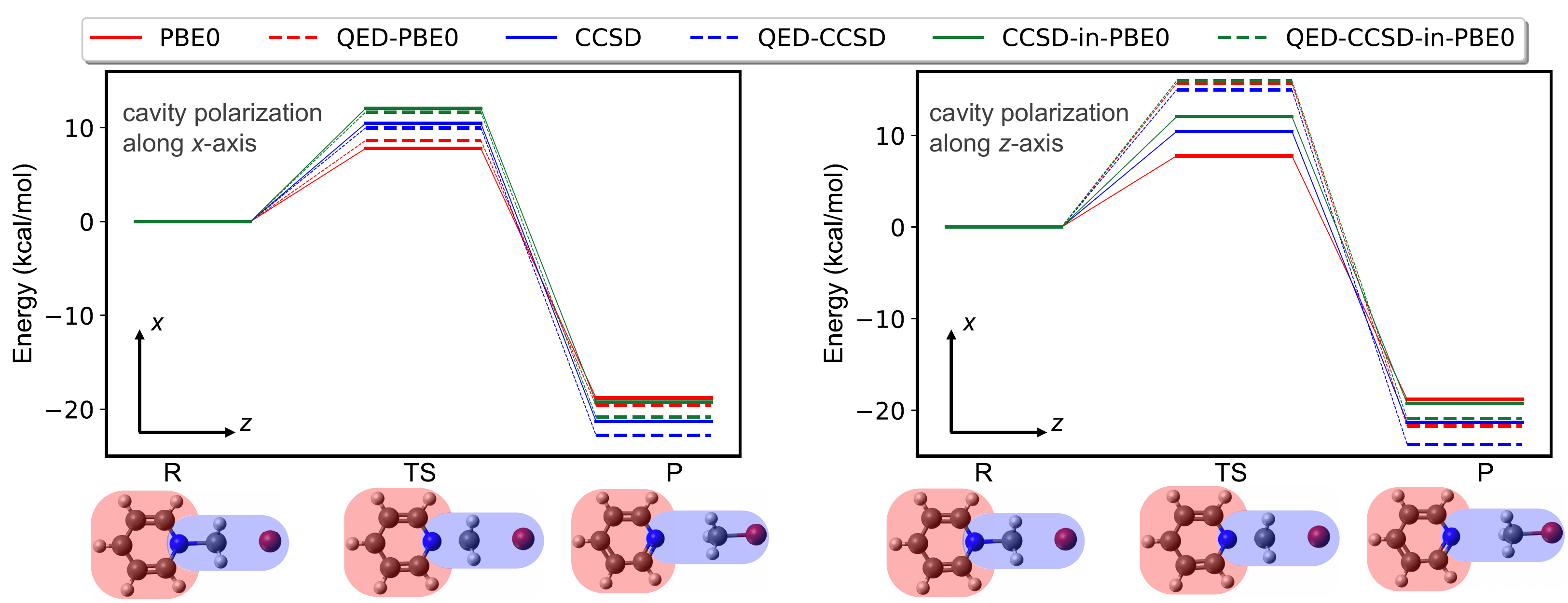}
  \caption{Reaction diagram for methyl transfer in pyridine with methyl bromide calculated with PBE0 (red), CCSD (blue), and CCSD-in-PBE0 (green) outside (solid) and inside (dashed) an optical cavity utilizing the 6-31G(d) basis set. The QED calculations employ $\omega_\text{cav}=3$ eV and $\lambda=0.1$ a.u. cavity parameters with the photon mode polarized along the $x$ (left panel), and $z$ (right panel) directions. The images of reactant (R), transition state (TS), and product (P) structures also depict the subsystem A in blue (treated with CCSD/QED-CCSD) and subsystem B in red (treated with PBE0/QED-PBE0). }
  \label{fig:sn2}
\end{figure*}

In our first example, we study the effect of optical cavity on the Menshutkin $\text{S}_\text{N}2$ reaction. Figure~\ref{fig:sn2} shows the reaction energy diagram of the nucleophilic methyl transfer process in pyridine with methyl bromide calculated with the PBE0, CCSD, and CCSD-in-PBE0 methods (solid lines) along with their QED counterparts (dashed lines) employing the 6-31G(d) basis set.~\cite{ditchfield1971self,hariharan1973influence,rassolov20016} The QED calculations were performed with cavity mode polarized along the $x$ (left panel) and $z$ (right panel) directions and by employing $\omega_\text{cav}=3$ eV and $\lambda=0.1$ a.u. cavity parameters that are both within the range of current experimental setups~\cite{Eizner2019,wu2021} as discussed earlier in the text. Moreover, Fig.~\ref{fig:sn2} indicates partitioning of the full system into the subsystem A (blue domain) and the subsystem B (red domain). The subsystem A is treated with the QED-CCSD method and the subsystem B is treated with the QED-SCF method. The effects of the cavity on change in the reaction energies and barriers for the cavity polarized along the $x$, $y$, and $z$ directions calculated with the QED-HF, QED-PBE, QED-PBE0, QED-B3LYP, QED-CCSD, QED-CCSD-in-HF, QED-CCSD-in-PBE, QED-CCSD-in-PBE0, and QED-CCSD-in-B3LYP methods using the 6-31G(d) basis set~\cite{ditchfield1971self,hariharan1973influence,rassolov20016} are provided in Table \ref{table:reaction_profile_SN2}. This Table shows that all the embedding methods are in excellent agreement with the QED-CCSD method, and that they are able to accurately describe changes in the energy reactions and barriers due to cavity.  

\begin{table}[htbp]
\caption{Change in the reaction energy barrier (TS)\textsuperscript{a} and reaction energy ($\Delta E$)\textsuperscript{b} (in kcal/mol) for methyl transfer in pyridine with methyl bromide inside an optical cavity.}
\centering
\begin{tabular}{c | c c | c c | c c}
\hline
 method & \multicolumn{2}{c|}{$x$ direction} & \multicolumn{2}{c|}{$y$ direction} & \multicolumn{2}{c}{$z$ direction} \\
 \hline\hline
 &      TS    &     $\Delta E$     &     TS      &     $\Delta E$     &     TS      &      $\Delta E$    \\
  \hline
QED-HF           &  0.39 & -0.84 & -0.22 & -1.88 & 5.00 &  1.01\\
QED-PBE          &  1.63 &  0.38 & -0.18 & -1.75 & 5.38 & -7.85\\
QED-PBE0         &  0.86 & -0.51 & -0.38 & -2.01 & 8.07 & -2.38\\
QED-B3LYP        &  1.06 & -0.30 & -0.35 & -1.94 & 7.50 & -3.10\\
QED-CCSD         & -0.50 & -1.55 & -0.38 & -1.38 & 3.45 & -2.41\\
QED-CCSD-in-HF   & -0.22 & -1.34 & -0.42 & -1.70 & 3.69 & -2.09\\
QED-CCSD-in-PBE  & -0.65 & -1.89 & -0.45 & -1.72 & 4.67 & -1.19\\
QED-CCSD-in-PBE0 & -0.42 & -1.63 & -0.41 & -1.66 & 4.05 & -1.88\\
QED-CCSD-in-B3LYP& -0.47 & -1.66 & -0.43 & -1.70 & 4.09 & -1.73\\
\hline
\end{tabular}
\textsuperscript{a}\small Effect of the cavity on the reaction energy barrier is calculated as the difference between the reaction energy barrier obtained with the QED method and the corresponding conventional electronic structure method. \\

\textsuperscript{b}\small Effect of the cavity on the on the reaction energy (i.e., the difference between the energies of the product and reactant) is calculated as the difference between the reaction energy obtained with the QED method and the corresponding conventional electronic structure method.\\

\label{table:reaction_profile_SN2}
\end{table}

As shown in Fig. \ref{fig:sn2}, the reaction barrier for the methyl transfer depends on the choice of the electronic structure method, where PBE0 underestimates the reaction barrier by $\sim$3 kcal/mol relative to the CCSD method. The embedding CCSD-in-PBE0 method shows better agreement relative to CCSD, where the reaction barrier is overestimated by only $\sim$1 kcal/mol. We note that extension of the embedding domain to include two adjacent CH groups will reduce this discrepancy to $\sim$0.2 kcal/mol. More information about performance of different SCF and CCSD-in-SCF methods, as well as different embedding domains is provided in Table S1 of the supplementary material.
The QED-HF, QED-PBE, QED-PBE0, and QED-B3LYP methods predicts increase in reaction barrier when the cavity mode is polarized along the $x$ direction, which is in stark contrast relative to the QED-CCSD method that predicts decrease in reaction barrier. This discrepancy is due to lack of the electron-photon correlation effects in the QED-SCF methods. All of the embedding QED-CCSD-in-HF, QED-CCSD-in-PBE, QED-CCSD-in-PBE0, and QED-CCSD-in-B3LYP methods predicts the same qualitative trend as the QED-CCSD method. This important finding indicates the local nature of the strong light-matter interaction and that relatively small region can be treated at the high level for qualitatively and quantitatively correct description of the cavity effect. For the cavity mode with polarization along the $y$ direction, all the QED methods predicts a decrease in reaction barrier and they are all in agreement with each other. Lastly, in the case of the cavity mode polarized in the $z$ direction, all studied QED methods predicts an increase of reaction barrier. The greatest increase in the reaction barrier is observed with the QED-SCF methods. Inclusion of the correlation effects between electrons and photons with either the QED-CCSD method or the embedding methods reduces this value by a few kcal/mol. 

Next, we discuss the cavity effect on the reaction energy for the same Menshutkin reaction, which is calculated as the energy difference between the product and the reactant. In the case of cavity mode with polarization along the $x$ direction, all of the QED-SCF (i.e. QED-HF, QED-PBE, QED-PBE0, and QED-B3LYP) methods underestimate the effect of the cavity on reaction energy relative to the QED-CCSD method, whereas the embedding methods are in an excellent agreement with the full QED-CCSD method. For the cavity mode with polarization along the $y$ direction, all the QED methods are in agreement with each other. Finally, in the case of cavity with the cavity mode polarized in the $z$ direction, all studied QED methods predict a decrease of the reaction energy in the presence of the optical cavity, whereas the QED-HF method predicts the opposite trend.


\begin{figure}[ht]
  \centering
  \includegraphics[width=3.25in]{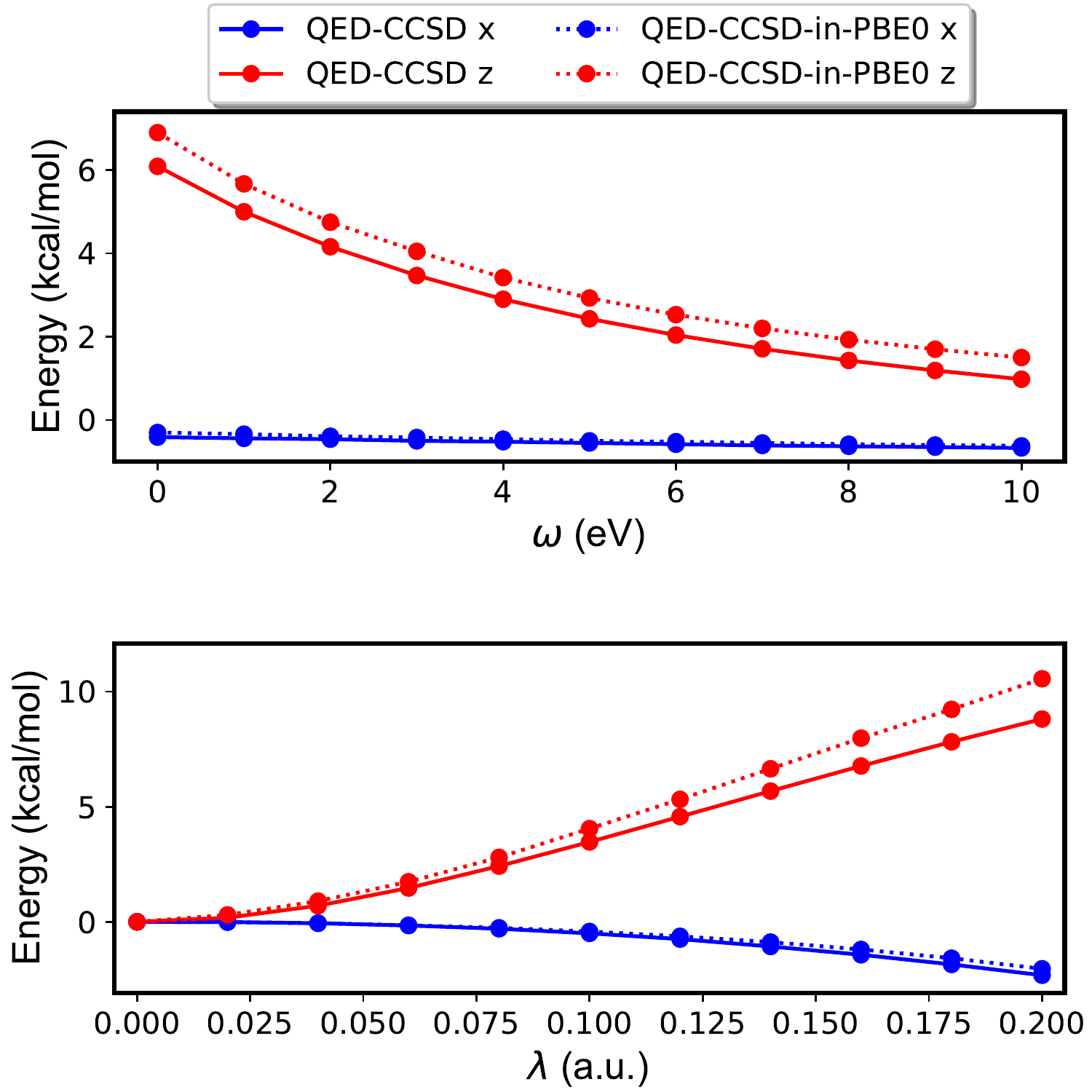}
  \caption{Change of the reaction barrier for the Menshutkin reaction as a function of cavity frequency (upper panels) and cavity coupling strength (lower panels) inside of optical cavity with light polarized along the $x$ (blue) and $z$ (red) directions. Solid lines correspond to the QED-CCSD calculations, whereas dotted lines correspond to the QED-CCSD-in-PBE0 calculations. The upper panel is calculated with the coupling strength of 0.1 a.u., whereas the lower panel is calculated with the cavity frequency of 3 eV.}
  \label{fig:barrier_vs_omega_and_lambda}
\end{figure}

The upper panel of Fig.~\ref{fig:barrier_vs_omega_and_lambda} shows the effect of cavity on reaction barrier for the Menshutkin reaction as the cavity frequency increases from 0 eV to 10 eV, whereas the lower panel shows the effect of cavity of the same reaction as the cavity coupling strength increases from 0 a.u. to 0.2 a.u. The upper and lower panels are calculated for the cavity coupling strength of 0.1 a.u. and for the cavity frequency of 3 eV, respectively. The changes in barriers are calculated with the QED-CCSD method (solid lines) and QED-CCSD-in-PBE0 method (dotted lines) in a cavity with the mode polarized along $x$ (blue lines) and $z$ (red lines) directions. As indicated in Fig.~\ref{fig:barrier_vs_omega_and_lambda}, the embedding QED-CCSD-in-PBE0 method is in an excellent agreement with the QED-CCSD method for this ranges of cavity coupling strengths and cavity frequencies. The upper panel also shows that in the case of the cavity mode polarized along the $x$ and $z$ directions the reaction barriers are decreasing as the cavity frequency increases. We note that the QED-SCF methods do not have dependence on cavity frequency as discussed in Ref.~\citenum{haugland2020coupled} and ~\citenum{deprince2021cavity}. The lower panel shows that in the case of very large values of coupling strength, the reaction barrier decreases by $\sim$2 kcal/mol when cavity mode is polarized along the $x$ direction, whereas when cavity mode is polarized along the $z$ direction the reaction barrier barrier increases by $\sim$10 kcal/mol. 

\begin{figure*}[ht!]
  \centering
  \includegraphics[width=6.5in]{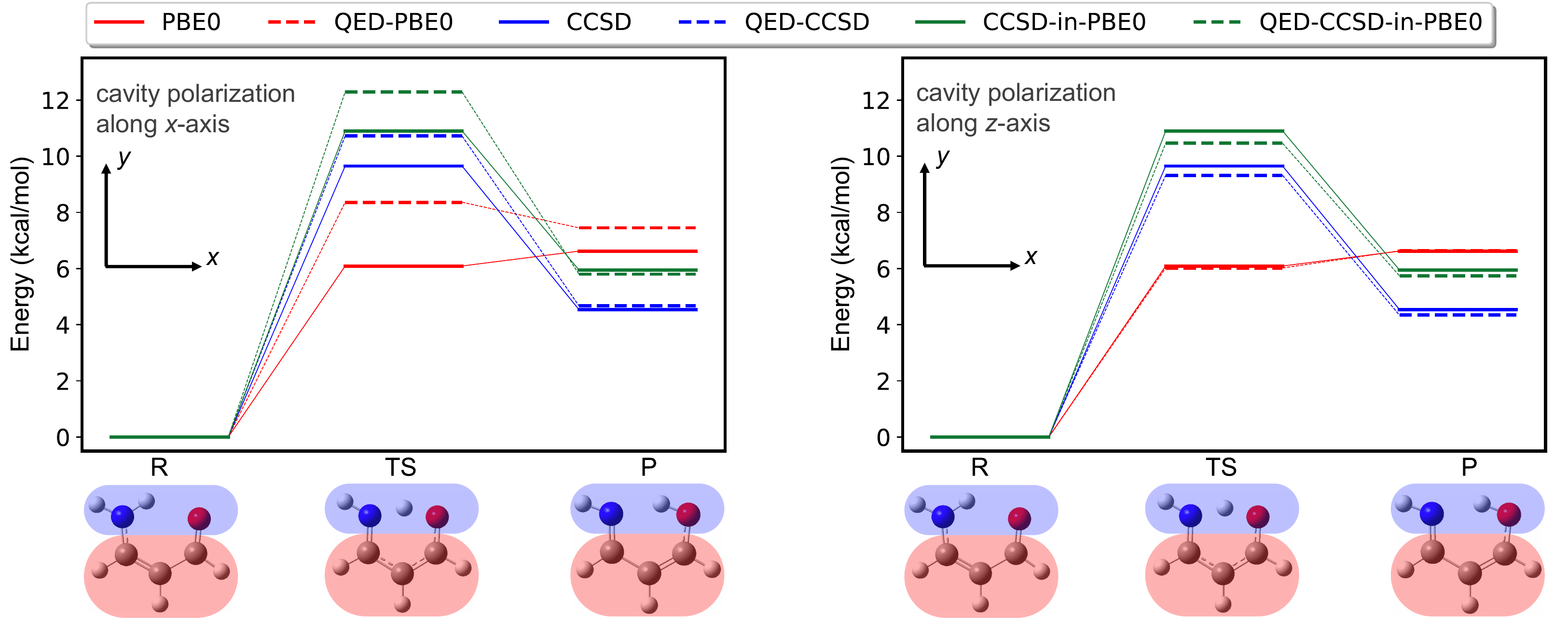}
  \caption{Reaction diagram for proton transfer in the aminopropenal molecule calculated with PBE0 (red), CCSD (blue), and CCSD-in-PBE0 (green) outside (solid) and inside (dashed) an optical cavity utilizing the cc-pVDZ basis set. The QED calculations employ $\omega_\text{cav}=3$ eV and $\lambda=0.1$ a.u. cavity parameters with the photon mode polarized along the $x$ (left panel), and $z$ (right panel) directions. The images of reactant (R), transition state (TS), and product (P) structures also depict the subsystem A in blue (treated with CCSD/QED-CCSD) and the subsystem B in red (treated with PBE0/QED-PBE0).}
  \label{fig:aminopropenal}
\end{figure*}

\begin{table}[htbp]
\caption{Change in the reaction energy barrier (TS)\textsuperscript{a} and reaction energy ($\Delta E$)\textsuperscript{b} (in kcal/mol) for proton transfer in aminopropenal inside an optical cavity.}
\centering
\begin{tabular}{c | c c | c c | c c}
\hline
 method & \multicolumn{2}{c|}{$x$ direction} & \multicolumn{2}{c|}{$y$ direction} & \multicolumn{2}{c}{$z$ direction} \\
 \hline\hline
 &      TS    &     $\Delta E$     &     TS      &     $\Delta E$     &     TS      &      $\Delta E$    \\
  \hline
QED-HF           &  2.00 &  0.94 &  0.59 &  0.37 & -0.01 &  0.06\\
QED-PBE          &  2.21 &  0.81 &  0.15 & -0.18 & -0.10 &  0.00\\
QED-PBE0         &  2.27 &  0.83 &  0.36 & -0.03 & -0.07 &  0.01\\
QED-B3LYP        &  2.26 &  0.82 &  0.32 & -0.07 & -0.07 &  0.01\\
QED-CCSD         &  1.07 &  0.13 &  0.06 & -0.22 & -0.33 & -0.20\\
QED-CCSD-in-HF   &  1.32 &  0.20 &  0.00 & -0.32 & -0.27 & -0.23\\
QED-CCSD-in-PBE  &  1.72 & -0.31 & -0.02 & -0.30 & -0.29 & -0.20\\
QED-CCSD-in-PBE0 &  1.55 & -0.15 & -0.03 & -0.32 & -0.28 & -0.21\\
QED-CCSD-in-B3LYP&  1.64 & -0.21 & -0.03 & -0.34 & -0.28 & -0.22\\
\hline
\end{tabular}
\textsuperscript{a}\small Effect of the cavity on the reaction energy barrier is calculated as the difference between the reaction energy barrier obtained with the QED method and the corresponding conventional electronic structure method. \\

\textsuperscript{b}\small Effect of the cavity on the on the reaction energy (i.e., the difference between the energies of the product and reactant) is calculated as the difference between the reaction energy obtained with the QED method and the corresponding conventional electronic structure method.\\

\label{table:reaction_profile_aminopropenal}
\end{table}

As our next example, we study the effect of optical cavity on the proton transfer in the aminopropenal molecule. Figure~\ref{fig:aminopropenal} shows the reaction energy diagram for proton transfer reaction in the aminopropenal molecule calculated with the PBE0, CCSD, and CCSD-in-PBE0 methods outside cavity (solid lines) and inside cavity (dashed lines) employing the cc-pVDZ basis set.~\cite{dunning1989gaussian} The QED calculations are performed with cavity parameters $\omega_\text{cav}=3$ eV and coupling strength $\lambda=0.1$ a.u. with light polarized along the $x$ (left) and $z$ (right) directions. Both left and right panels include geometries of reactant (R), transition state (TS), and product (P) along with partitioning of the full system into the subsystem A (blue region) and the subsystem B (red region).  Table~\ref{table:reaction_profile_aminopropenal} provides the changes in reaction energies and barriers calculated with QED-HF, QED-PBE, QED-PBE0, QED-B3LYP, QED-CCSD, QED-CCSD-in-HF, QED-CCSD-in-PBE, QED-CCSD-in-PBE0, and QE-CCSD-in-B3LYP methods with the cavity mode polarized along all three directions. As it was the case for Menshutkin reaction, the reaction energy profile greatly depends on the choice of electronic structure method as indicated in Fig.~\ref{fig:aminopropenal} and Table S2 of the supplementary material. Interestingly, both PBE and PBE0 predict the barrier-less proton transfer reaction using these geometries, whereas the HF, B3LYP, CCSD, and all of the CCSD-in-SCF methods predict the process with the barrier. The HF method greatly overestimates the reaction barrier relative to the CCSD method, whereas the CCSD-in-PBE0 method predicts the barrier that is $\sim$1.5~kcal/mol of the CCSD barrier. Furthermore, all of the CCSD-in-SCF methods predict the reaction energy that is in a good agreement with the CCSD predictions.

For the cavity with the mode polarized along the $x$ direction, the greatest increase in the barrier due to cavity is observed for the QED-SCF methods. Inclusion of the electron-photon correlation effects with the QED-CCSD method reduces this change by $\sim$1 kcal/mol. All the embedding QED-CCSD-in-SCF methods show an increase of the barrier that is in between the one observed with the QED-CCSD and QED-SCF methods. For the cavity mode polarized along the $y$ direction, the QED-CCSD and QED-CCSD-in-SCF methods show very small change ($<0.1$~kcal/mol) in barrier due to cavity, whereas the QED-SCF methods predicts more pronounced increase in the reaction barrier. When the cavity mode is polarized along the $z$ direction, all of the QED-SCF methods show a negligible decrease of the reaction barrier, whereas upon inclusion of the correlation effects between electrons and photons with the QED-CCSD and QED-CCSD-in-SCF methods, this change becomes more pronounced. Importantly, all the QED-embedding methods are in an excellent agreement with the QED-CCSD method.  

Next, we discuss the effect of the cavity on the reaction energies for the same proton transfer reaction. In the case when the cavity mode is polarized along the $x$ direction, the QED-SCF methods predict significant increase of the reaction energy due to cavity, whereas the QED-CCSD method predicts a more modest change. While QED-CCSD-in-HF is in a good agreement with the QED-CCSD method, the QED-CCSD-in-PBE, QED-CCSD-in-PBE0, and QED-CCSD-in-B3LYP on the other hand predict completely opposite trend. For the cavity with the mode polarized along the $y$ direction, all of the QED-SCF methods, except for the QED-HF method, predict small decrease in reaction energies, whereas for the QED-CCSD and QED-CCSD-in-SCF methods this decrease is slightly more pronounced. Finally, for the cavity mode polarized along the $z$ direction, all of the QED-SCF methods show small increase in the reaction energy due to cavity, whereas the QED-CCSD and QED-CCSD-in-SCF methods predict the opposite trend. 

\begin{figure}[ht]
  \centering
  \includegraphics[width=3.25in]{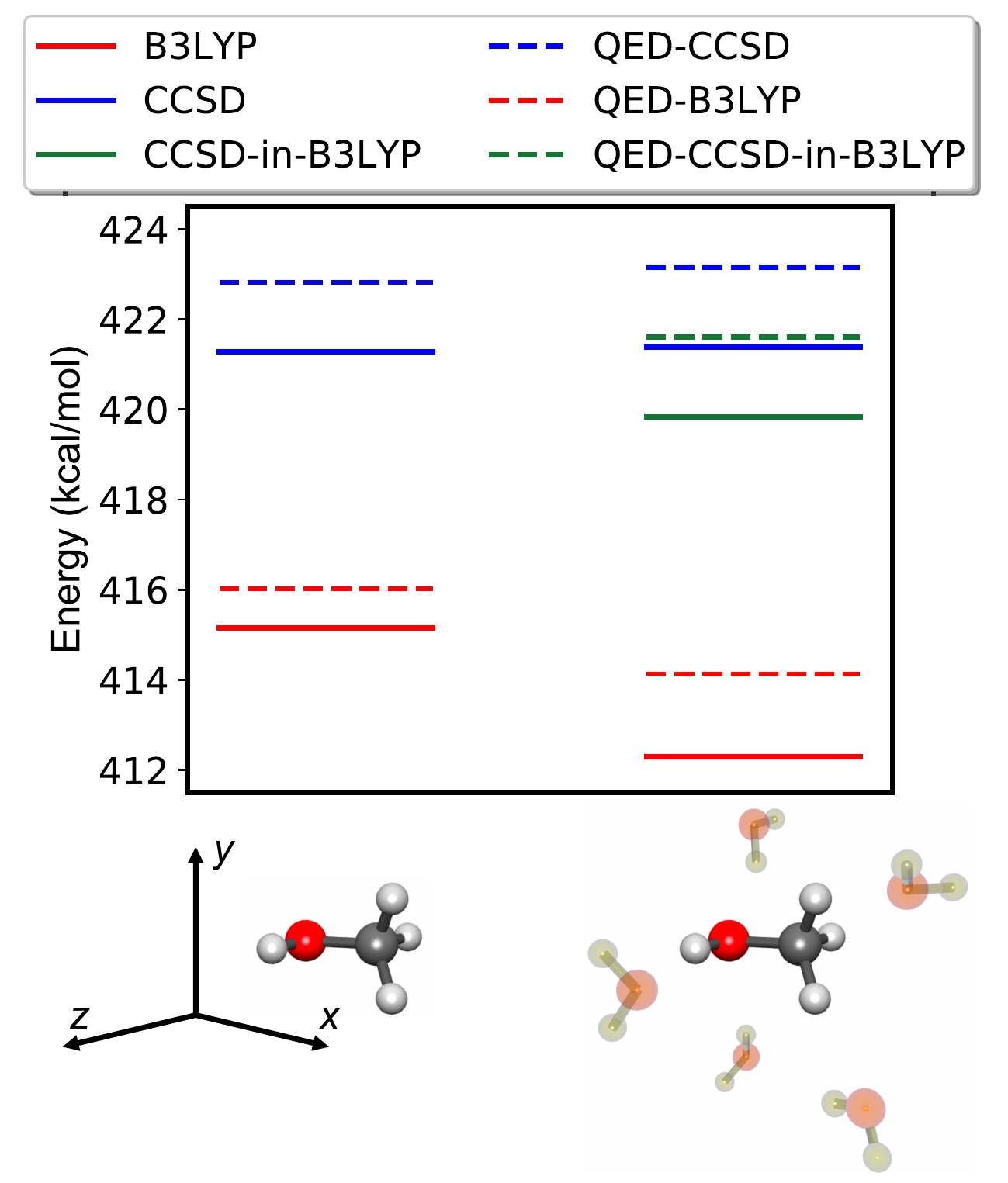}
  \caption{Proton binding energies (PBEn) for methanol in the gas phase (left) and explicit solvent (right) calculated with B3LYP, CCSD, and CCSD-in-B3LYP methods outside (solid lines) and inside (dashed lines) optical cavity employing the cc-pVDZ basis set. The QED calculations employ $\omega_\text{cav}=3$ eV and $\lambda=0.1$ a.u. cavity parameters with the photon mode polarized along the $z$ direction (along dissociating O-H bond). The image of methanol in explicit water are also shown (right).}
  \label{fig:PBE}
\end{figure}

\begin{table}[htbp]
\caption{Proton binding energies (in kcal/mol) for methanol in the gas phase and in explicit solvent calculated outside and inside an optical cavity.}
\centering
\begin{tabular}{c | c c | c c }
\hline
 method & \multicolumn{2}{c|}{   MeOH    } & \multicolumn{2}{c}{MeOH-in-5H$_2$O} \\
 \hline\hline
 &       outside    &      inside     &      outside      &     inside\\
 &       cavity    &      cavity     &      cavity      &     cavity\\
  \hline
QED-HF           & 419.50 & 420.02 & 421.43 & 422.38 \\
QED-PBE          & 412.70 & 413.81 & 407.63 & 410.18 \\
QED-PBE0         & 415.64 & 416.50 & 413.85 & 415.47 \\
QED-B3LYP        & 415.15 & 416.02 & 412.30 & 414.13 \\
QED-CCSD         & 421.28 & 422.82 & 421.38 & 423.15 \\
QED-CCSD-in-HF   & -      & -      & 421.34 & 423.17 \\
QED-CCSD-in-PBE  & -      & -      & 419.53 & 421.28 \\
QED-CCSD-in-PBE0 & -      & -      & 419.65 & 421.47 \\
QED-CCSD-in-B3LYP& -      & -      & 419.83 & 421.60 \\
\hline
\end{tabular}


\label{table:PBE}
\end{table}

In our last example, we investigate the cavity effect on the proton binding energy (PBEn) for methanol (MeOH) with explicit water solvent. The PBEn is calculated as the energy difference between MeO$^-$ and MeOH. Figure~\ref{fig:PBE} depicts the PBEn for MeOH molecule in the gas phase (left side of Fig.~\ref{fig:PBE}) and in explicit solvent consisting of five water molecules (right side of Fig.~\ref{fig:PBE}) calculated with the B3LYP, CCSD, and QED-CCSD-in-B3LYP methods outside (solid lines) and inside (dashed lines) the cavity employing the cc-pVDZ basis set.~\cite{dunning1989gaussian} The QED calculations were performed with the cavity mode polarized along the $z$ direction (along dissociating O-H bond) with the cavity parameters $\omega_\text{cav}=3$ eV and $\lambda=0.1$ a.u. Moreover, Fig.~\ref{fig:PBE} also contains the coordinate frame along with the MeOH molecule in the gas phase (left) and in an explicit solvent (right). Numerical values of the PBEn calculated with different SCF and CCSD-in-SCF methods outside and inside cavity are given in Table \ref{table:PBE}. It is indicative that for the investigated system in the gas phase all of the studied QED-SCF methods underestimate the effect of cavity relative to the QED-CCSD predictions. Moreover, the QED-SCF methods predicts two fold increase of the PBEn due to cavity upon inclusion of the solvent, whereas this change for the QED-CCSD method remains nearly constant (1.54 kcal/mol vs. 1.76 kcal/mol). This is primarily due to an inadequate treatment of the dipole self energy with the QED-SCF methods. The embedding methods, where only MeOH is treated with the QED-CCSD method, are in excellent agreement with the QED-CCSD method for the full system.

\section{Conclusions}

In this work, we have developed and implemented the QED-CCSD-in-SCF embedding method for polaritonic chemistry in which only the chemically important region is treated with the accurate but computationally expensive QED-CCSD method, whereas the environment is treated at the computationally more efficient QED-SCF level of the theory. We illustrate the performance of the method by studying the effect of cavity on the methyl and proton transfer reactions, as well as protonation reaction in an explicit solvent. The results obtained with the embedding method are in excellent agreement with results obtained using a more expensive QED-CCSD method. Moreover, we observe ten-fold computational speed-up of the QED-CCSD-in-SCF vs. QED-CCSD method for explored systems. Out of different studied QED-CCSD-in-SCF methods, the QED-CCSD-in-HF shows the best performance. We show that the correlation effects between the quantum particles is crucial for an accurate description of the effect of the optical cavity. We further show that the strong light-matter coupling is relatively local in nature and only a small chemically important region has to be treated with the correlated QED-CCSD method for achieving a reliable accuracy. The development and analysis presented in this work will serve as a guideline for development of novel polaritonic quantum chemistry methods and it provides a valuable insights into polaritonic systems inside complex environments. 

\section*{Supplementary Material}
See the supplementary material for the reaction energies and barriers of the Menshutkin reaction and proton transfer reaction, and optimized structures.

\begin{acknowledgments}
We acknowledge financial support from the Cluster of Excellence 'CUI: Advanced Imaging of Matter' of the Deutsche Forschungsgemeinschaft (DFG) - EXC 2056 - project ID 390715994. We also acknowledge support from the Max Planck–New York Center for Non-Equilibrium Quantum Phenomena. 
The Flatiron Institute is a division of the Simons Foundation.
\end{acknowledgments}

\section*{AUTHOR DECLARATIONS}

\textbf{Conflict of Interest}\\ 

The authors have no conflicts of interest to disclose.



\section*{Data Availability Statement}

The data that support the findings of this study are available within this article and its supplementary material.

\bibliography{Journal_Short_Name,references}

\begin{thebibliography}{66}%
\makeatletter
\providecommand \@ifxundefined [1]{%
 \@ifx{#1\undefined}
}%
\providecommand \@ifnum [1]{%
 \ifnum #1\expandafter \@firstoftwo
 \else \expandafter \@secondoftwo
 \fi
}%
\providecommand \@ifx [1]{%
 \ifx #1\expandafter \@firstoftwo
 \else \expandafter \@secondoftwo
 \fi
}%
\providecommand \natexlab [1]{#1}%
\providecommand \enquote  [1]{``#1''}%
\providecommand \bibnamefont  [1]{#1}%
\providecommand \bibfnamefont [1]{#1}%
\providecommand \citenamefont [1]{#1}%
\providecommand \href@noop [0]{\@secondoftwo}%
\providecommand \href [0]{\begingroup \@sanitize@url \@href}%
\providecommand \@href[1]{\@@startlink{#1}\@@href}%
\providecommand \@@href[1]{\endgroup#1\@@endlink}%
\providecommand \@sanitize@url [0]{\catcode `\\12\catcode `\$12\catcode
  `\&12\catcode `\#12\catcode `\^12\catcode `\_12\catcode `\%12\relax}%
\providecommand \@@startlink[1]{}%
\providecommand \@@endlink[0]{}%
\providecommand \url  [0]{\begingroup\@sanitize@url \@url }%
\providecommand \@url [1]{\endgroup\@href {#1}{\urlprefix }}%
\providecommand \urlprefix  [0]{URL }%
\providecommand \Eprint [0]{\href }%
\providecommand \doibase [0]{http://dx.doi.org/}%
\providecommand \selectlanguage [0]{\@gobble}%
\providecommand \bibinfo  [0]{\@secondoftwo}%
\providecommand \bibfield  [0]{\@secondoftwo}%
\providecommand \translation [1]{[#1]}%
\providecommand \BibitemOpen [0]{}%
\providecommand \bibitemStop [0]{}%
\providecommand \bibitemNoStop [0]{.\EOS\space}%
\providecommand \EOS [0]{\spacefactor3000\relax}%
\providecommand \BibitemShut  [1]{\csname bibitem#1\endcsname}%
\let\auto@bib@innerbib\@empty
\bibitem [{\citenamefont {Lather}\ \emph {et~al.}(2019)\citenamefont {Lather},
  \citenamefont {Bhatt}, \citenamefont {Thomas}, \citenamefont {Ebbesen},\ and\
  \citenamefont {George}}]{lather2019cavity}%
  \BibitemOpen
  \bibfield  {author} {\bibinfo {author} {\bibfnamefont {J.}~\bibnamefont
  {Lather}}, \bibinfo {author} {\bibfnamefont {P.}~\bibnamefont {Bhatt}},
  \bibinfo {author} {\bibfnamefont {A.}~\bibnamefont {Thomas}}, \bibinfo
  {author} {\bibfnamefont {T.~W.}\ \bibnamefont {Ebbesen}}, \ and\ \bibinfo
  {author} {\bibfnamefont {J.}~\bibnamefont {George}},\ }\bibfield  {title}
  {\enquote {\bibinfo {title} {Cavity catalysis by cooperative vibrational
  strong coupling of reactant and solvent molecules},}\ }\href@noop {}
  {\bibfield  {journal} {\bibinfo  {journal} {Angew. Chem.}\ }\textbf {\bibinfo
  {volume} {58}},\ \bibinfo {pages} {10635--10638} (\bibinfo {year}
  {2019})}\BibitemShut {NoStop}%
\bibitem [{\citenamefont {Campos-Gonzalez-Angulo}, \citenamefont {Ribeiro},\
  and\ \citenamefont {Yuen-Zhou}(2019)}]{campos2019resonant}%
  \BibitemOpen
  \bibfield  {author} {\bibinfo {author} {\bibfnamefont {J.~A.}\ \bibnamefont
  {Campos-Gonzalez-Angulo}}, \bibinfo {author} {\bibfnamefont {R.~F.}\
  \bibnamefont {Ribeiro}}, \ and\ \bibinfo {author} {\bibfnamefont
  {J.}~\bibnamefont {Yuen-Zhou}},\ }\bibfield  {title} {\enquote {\bibinfo
  {title} {Resonant catalysis of thermally activated chemical reactions with
  vibrational polaritons},}\ }\href@noop {} {\bibfield  {journal} {\bibinfo
  {journal} {Nat. Commun.}\ }\textbf {\bibinfo {volume} {10}},\ \bibinfo
  {pages} {1--8} (\bibinfo {year} {2019})}\BibitemShut {NoStop}%
\bibitem [{\citenamefont {Climent}\ \emph {et~al.}(2019)\citenamefont
  {Climent}, \citenamefont {Galego}, \citenamefont {Garcia-Vidal},\ and\
  \citenamefont {Feist}}]{climent2019plasmonic}%
  \BibitemOpen
  \bibfield  {author} {\bibinfo {author} {\bibfnamefont {C.}~\bibnamefont
  {Climent}}, \bibinfo {author} {\bibfnamefont {J.}~\bibnamefont {Galego}},
  \bibinfo {author} {\bibfnamefont {F.~J.}\ \bibnamefont {Garcia-Vidal}}, \
  and\ \bibinfo {author} {\bibfnamefont {J.}~\bibnamefont {Feist}},\ }\bibfield
   {title} {\enquote {\bibinfo {title} {Plasmonic nanocavities enable
  self-induced electrostatic catalysis},}\ }\href@noop {} {\bibfield  {journal}
  {\bibinfo  {journal} {Angew. Chem. Int. Ed.}\ }\textbf {\bibinfo {volume}
  {58}},\ \bibinfo {pages} {8698--8702} (\bibinfo {year} {2019})}\BibitemShut
  {NoStop}%
\bibitem [{\citenamefont {Thomas}\ \emph {et~al.}(2016)\citenamefont {Thomas},
  \citenamefont {George}, \citenamefont {Shalabney}, \citenamefont {Dryzhakov},
  \citenamefont {Varma}, \citenamefont {Moran}, \citenamefont {Chervy},
  \citenamefont {Zhong}, \citenamefont {Devaux}, \citenamefont {Genet} \emph
  {et~al.}}]{thomas2016ground}%
  \BibitemOpen
  \bibfield  {author} {\bibinfo {author} {\bibfnamefont {A.}~\bibnamefont
  {Thomas}}, \bibinfo {author} {\bibfnamefont {J.}~\bibnamefont {George}},
  \bibinfo {author} {\bibfnamefont {A.}~\bibnamefont {Shalabney}}, \bibinfo
  {author} {\bibfnamefont {M.}~\bibnamefont {Dryzhakov}}, \bibinfo {author}
  {\bibfnamefont {S.~J.}\ \bibnamefont {Varma}}, \bibinfo {author}
  {\bibfnamefont {J.}~\bibnamefont {Moran}}, \bibinfo {author} {\bibfnamefont
  {T.}~\bibnamefont {Chervy}}, \bibinfo {author} {\bibfnamefont
  {X.}~\bibnamefont {Zhong}}, \bibinfo {author} {\bibfnamefont
  {E.}~\bibnamefont {Devaux}}, \bibinfo {author} {\bibfnamefont
  {C.}~\bibnamefont {Genet}},  \emph {et~al.},\ }\bibfield  {title} {\enquote
  {\bibinfo {title} {Ground-state chemical reactivity under vibrational
  coupling to the vacuum electromagnetic field},}\ }\href@noop {} {\bibfield
  {journal} {\bibinfo  {journal} {Angew. Chem.}\ }\textbf {\bibinfo {volume}
  {128}},\ \bibinfo {pages} {11634--11638} (\bibinfo {year}
  {2016})}\BibitemShut {NoStop}%
\bibitem [{\citenamefont {Hutchison}\ \emph {et~al.}(2012)\citenamefont
  {Hutchison}, \citenamefont {Schwartz}, \citenamefont {Genet}, \citenamefont
  {Devaux},\ and\ \citenamefont {Ebbesen}}]{hutchison2012modifying}%
  \BibitemOpen
  \bibfield  {author} {\bibinfo {author} {\bibfnamefont {J.~A.}\ \bibnamefont
  {Hutchison}}, \bibinfo {author} {\bibfnamefont {T.}~\bibnamefont {Schwartz}},
  \bibinfo {author} {\bibfnamefont {C.}~\bibnamefont {Genet}}, \bibinfo
  {author} {\bibfnamefont {E.}~\bibnamefont {Devaux}}, \ and\ \bibinfo {author}
  {\bibfnamefont {T.~W.}\ \bibnamefont {Ebbesen}},\ }\bibfield  {title}
  {\enquote {\bibinfo {title} {Modifying chemical landscapes by coupling to
  vacuum fields},}\ }\href@noop {} {\bibfield  {journal} {\bibinfo  {journal}
  {Angew. Chem.}\ }\textbf {\bibinfo {volume} {51}},\ \bibinfo {pages}
  {1592--1596} (\bibinfo {year} {2012})}\BibitemShut {NoStop}%
\bibitem [{\citenamefont {Fregoni}\ \emph {et~al.}(2018)\citenamefont
  {Fregoni}, \citenamefont {Granucci}, \citenamefont {Coccia}, \citenamefont
  {Persico},\ and\ \citenamefont {Corni}}]{fregoni2018manipulating}%
  \BibitemOpen
  \bibfield  {author} {\bibinfo {author} {\bibfnamefont {J.}~\bibnamefont
  {Fregoni}}, \bibinfo {author} {\bibfnamefont {G.}~\bibnamefont {Granucci}},
  \bibinfo {author} {\bibfnamefont {E.}~\bibnamefont {Coccia}}, \bibinfo
  {author} {\bibfnamefont {M.}~\bibnamefont {Persico}}, \ and\ \bibinfo
  {author} {\bibfnamefont {S.}~\bibnamefont {Corni}},\ }\bibfield  {title}
  {\enquote {\bibinfo {title} {Manipulating azobenzene photoisomerization
  through strong light--molecule coupling},}\ }\href@noop {} {\bibfield
  {journal} {\bibinfo  {journal} {Nat. Chem.}\ }\textbf {\bibinfo {volume}
  {9}},\ \bibinfo {pages} {1--9} (\bibinfo {year} {2018})}\BibitemShut
  {NoStop}%
\bibitem [{\citenamefont {Thomas}\ \emph {et~al.}(2019)\citenamefont {Thomas},
  \citenamefont {Lethuillier-Karl}, \citenamefont {Nagarajan}, \citenamefont
  {Vergauwe}, \citenamefont {George}, \citenamefont {Chervy}, \citenamefont
  {Shalabney}, \citenamefont {Devaux}, \citenamefont {Genet}, \citenamefont
  {Moran} \emph {et~al.}}]{thomas2019tilting}%
  \BibitemOpen
  \bibfield  {author} {\bibinfo {author} {\bibfnamefont {A.}~\bibnamefont
  {Thomas}}, \bibinfo {author} {\bibfnamefont {L.}~\bibnamefont
  {Lethuillier-Karl}}, \bibinfo {author} {\bibfnamefont {K.}~\bibnamefont
  {Nagarajan}}, \bibinfo {author} {\bibfnamefont {R.~M.}\ \bibnamefont
  {Vergauwe}}, \bibinfo {author} {\bibfnamefont {J.}~\bibnamefont {George}},
  \bibinfo {author} {\bibfnamefont {T.}~\bibnamefont {Chervy}}, \bibinfo
  {author} {\bibfnamefont {A.}~\bibnamefont {Shalabney}}, \bibinfo {author}
  {\bibfnamefont {E.}~\bibnamefont {Devaux}}, \bibinfo {author} {\bibfnamefont
  {C.}~\bibnamefont {Genet}}, \bibinfo {author} {\bibfnamefont
  {J.}~\bibnamefont {Moran}},  \emph {et~al.},\ }\bibfield  {title} {\enquote
  {\bibinfo {title} {Tilting a ground-state reactivity landscape by vibrational
  strong coupling},}\ }\href@noop {} {\bibfield  {journal} {\bibinfo  {journal}
  {Science}\ }\textbf {\bibinfo {volume} {363}},\ \bibinfo {pages} {615--619}
  (\bibinfo {year} {2019})}\BibitemShut {NoStop}%
\bibitem [{\citenamefont {Ruggenthaler}\ \emph {et~al.}(2018)\citenamefont
  {Ruggenthaler}, \citenamefont {Tancogne-Dejean}, \citenamefont {Flick},
  \citenamefont {Appel},\ and\ \citenamefont
  {Rubio}}]{ruggenthaler2018quantum}%
  \BibitemOpen
  \bibfield  {author} {\bibinfo {author} {\bibfnamefont {M.}~\bibnamefont
  {Ruggenthaler}}, \bibinfo {author} {\bibfnamefont {N.}~\bibnamefont
  {Tancogne-Dejean}}, \bibinfo {author} {\bibfnamefont {J.}~\bibnamefont
  {Flick}}, \bibinfo {author} {\bibfnamefont {H.}~\bibnamefont {Appel}}, \ and\
  \bibinfo {author} {\bibfnamefont {A.}~\bibnamefont {Rubio}},\ }\bibfield
  {title} {\enquote {\bibinfo {title} {From a quantum-electrodynamical
  light-matter description to novel spectroscopies},}\ }\href@noop {}
  {\bibfield  {journal} {\bibinfo  {journal} {Nat. Rev. Chem.}\ }\textbf
  {\bibinfo {volume} {2}},\ \bibinfo {pages} {1--16} (\bibinfo {year}
  {2018})}\BibitemShut {NoStop}%
\bibitem [{\citenamefont {Lacombe}, \citenamefont {Hoffmann},\ and\
  \citenamefont {Maitra}(2019)}]{lacombe2019exact}%
  \BibitemOpen
  \bibfield  {author} {\bibinfo {author} {\bibfnamefont {L.}~\bibnamefont
  {Lacombe}}, \bibinfo {author} {\bibfnamefont {N.~M.}\ \bibnamefont
  {Hoffmann}}, \ and\ \bibinfo {author} {\bibfnamefont {N.~T.}\ \bibnamefont
  {Maitra}},\ }\bibfield  {title} {\enquote {\bibinfo {title} {Exact potential
  energy surface for molecules in cavities},}\ }\href@noop {} {\bibfield
  {journal} {\bibinfo  {journal} {Phys. Rev. Lett.}\ }\textbf {\bibinfo
  {volume} {123}},\ \bibinfo {pages} {083201} (\bibinfo {year}
  {2019})}\BibitemShut {NoStop}%
\bibitem [{\citenamefont {Li}, \citenamefont {Nitzan},\ and\ \citenamefont
  {Subotnik}(2021)}]{li2021collective}%
  \BibitemOpen
  \bibfield  {author} {\bibinfo {author} {\bibfnamefont {T.~E.}\ \bibnamefont
  {Li}}, \bibinfo {author} {\bibfnamefont {A.}~\bibnamefont {Nitzan}}, \ and\
  \bibinfo {author} {\bibfnamefont {J.~E.}\ \bibnamefont {Subotnik}},\
  }\bibfield  {title} {\enquote {\bibinfo {title} {Collective vibrational
  strong coupling effects on molecular vibrational relaxation and energy
  transfer: Numerical insights via cavity molecular dynamics simulations},}\
  }\href@noop {} {\bibfield  {journal} {\bibinfo  {journal} {Angew. Chem. Int.
  Ed.}\ }\textbf {\bibinfo {volume} {60}},\ \bibinfo {pages} {15533--15540}
  (\bibinfo {year} {2021})}\BibitemShut {NoStop}%
\bibitem [{\citenamefont {Sch{\"a}fer}\ \emph
  {et~al.}(2021{\natexlab{a}})\citenamefont {Sch{\"a}fer}, \citenamefont
  {Flick}, \citenamefont {Ronca}, \citenamefont {Narang},\ and\ \citenamefont
  {Rubio}}]{schafer2021shining}%
  \BibitemOpen
  \bibfield  {author} {\bibinfo {author} {\bibfnamefont {C.}~\bibnamefont
  {Sch{\"a}fer}}, \bibinfo {author} {\bibfnamefont {J.}~\bibnamefont {Flick}},
  \bibinfo {author} {\bibfnamefont {E.}~\bibnamefont {Ronca}}, \bibinfo
  {author} {\bibfnamefont {P.}~\bibnamefont {Narang}}, \ and\ \bibinfo {author}
  {\bibfnamefont {A.}~\bibnamefont {Rubio}},\ }\bibfield  {title} {\enquote
  {\bibinfo {title} {Shining light on the microscopic resonant mechanism
  responsible for cavity-mediated chemical reactivity},}\ }\href@noop {}
  {\bibfield  {journal} {\bibinfo  {journal} {arXiv preprint arXiv:2104.12429}\
  } (\bibinfo {year} {2021}{\natexlab{a}})}\BibitemShut {NoStop}%
\bibitem [{\citenamefont {Sidler}\ \emph {et~al.}(2020)\citenamefont {Sidler},
  \citenamefont {Ruggenthaler}, \citenamefont {Appel},\ and\ \citenamefont
  {Rubio}}]{sidler2020chemistry}%
  \BibitemOpen
  \bibfield  {author} {\bibinfo {author} {\bibfnamefont {D.}~\bibnamefont
  {Sidler}}, \bibinfo {author} {\bibfnamefont {M.}~\bibnamefont
  {Ruggenthaler}}, \bibinfo {author} {\bibfnamefont {H.}~\bibnamefont {Appel}},
  \ and\ \bibinfo {author} {\bibfnamefont {A.}~\bibnamefont {Rubio}},\
  }\bibfield  {title} {\enquote {\bibinfo {title} {Chemistry in quantum
  cavities: Exact results, the impact of thermal velocities, and modified
  dissociation},}\ }\href@noop {} {\bibfield  {journal} {\bibinfo  {journal}
  {J. Phys. Chem. Lett.}\ }\textbf {\bibinfo {volume} {11}},\ \bibinfo {pages}
  {7525--7530} (\bibinfo {year} {2020})}\BibitemShut {NoStop}%
\bibitem [{\citenamefont {Sidler}\ \emph {et~al.}(2021)\citenamefont {Sidler},
  \citenamefont {Ruggenthaler}, \citenamefont {Sch{\"a}fer}, \citenamefont
  {Ronca},\ and\ \citenamefont {Rubio}}]{sidler2021perspective}%
  \BibitemOpen
  \bibfield  {author} {\bibinfo {author} {\bibfnamefont {D.}~\bibnamefont
  {Sidler}}, \bibinfo {author} {\bibfnamefont {M.}~\bibnamefont
  {Ruggenthaler}}, \bibinfo {author} {\bibfnamefont {C.}~\bibnamefont
  {Sch{\"a}fer}}, \bibinfo {author} {\bibfnamefont {E.}~\bibnamefont {Ronca}},
  \ and\ \bibinfo {author} {\bibfnamefont {A.}~\bibnamefont {Rubio}},\
  }\bibfield  {title} {\enquote {\bibinfo {title} {A perspective on ab initio
  modeling of polaritonic chemistry: The role of non-equilibrium effects and
  quantum collectivity},}\ }\href@noop {} {\bibfield  {journal} {\bibinfo
  {journal} {arXiv preprint arXiv:2108.12244}\ } (\bibinfo {year}
  {2021})}\BibitemShut {NoStop}%
\bibitem [{\citenamefont {Ruggenthaler}\ \emph {et~al.}(2014)\citenamefont
  {Ruggenthaler}, \citenamefont {Flick}, \citenamefont {Pellegrini},
  \citenamefont {Appel}, \citenamefont {Tokatly},\ and\ \citenamefont
  {Rubio}}]{ruggenthaler2014quantum}%
  \BibitemOpen
  \bibfield  {author} {\bibinfo {author} {\bibfnamefont {M.}~\bibnamefont
  {Ruggenthaler}}, \bibinfo {author} {\bibfnamefont {J.}~\bibnamefont {Flick}},
  \bibinfo {author} {\bibfnamefont {C.}~\bibnamefont {Pellegrini}}, \bibinfo
  {author} {\bibfnamefont {H.}~\bibnamefont {Appel}}, \bibinfo {author}
  {\bibfnamefont {I.~V.}\ \bibnamefont {Tokatly}}, \ and\ \bibinfo {author}
  {\bibfnamefont {A.}~\bibnamefont {Rubio}},\ }\bibfield  {title} {\enquote
  {\bibinfo {title} {Quantum-electrodynamical density-functional theory:
  Bridging quantum optics and electronic-structure theory},}\ }\href@noop {}
  {\bibfield  {journal} {\bibinfo  {journal} {Phys. Rev. A}\ }\textbf {\bibinfo
  {volume} {90}},\ \bibinfo {pages} {012508} (\bibinfo {year}
  {2014})}\BibitemShut {NoStop}%
\bibitem [{\citenamefont {Flick}\ \emph {et~al.}(2015)\citenamefont {Flick},
  \citenamefont {Ruggenthaler}, \citenamefont {Appel},\ and\ \citenamefont
  {Rubio}}]{flick2015kohn}%
  \BibitemOpen
  \bibfield  {author} {\bibinfo {author} {\bibfnamefont {J.}~\bibnamefont
  {Flick}}, \bibinfo {author} {\bibfnamefont {M.}~\bibnamefont {Ruggenthaler}},
  \bibinfo {author} {\bibfnamefont {H.}~\bibnamefont {Appel}}, \ and\ \bibinfo
  {author} {\bibfnamefont {A.}~\bibnamefont {Rubio}},\ }\bibfield  {title}
  {\enquote {\bibinfo {title} {Kohn--sham approach to quantum electrodynamical
  density-functional theory: Exact time-dependent effective potentials in real
  space},}\ }\href@noop {} {\bibfield  {journal} {\bibinfo  {journal} {Proc.
  Natl. Acad. Sci. U.S.A.}\ }\textbf {\bibinfo {volume} {112}},\ \bibinfo
  {pages} {15285--15290} (\bibinfo {year} {2015})}\BibitemShut {NoStop}%
\bibitem [{\citenamefont {Flick}\ \emph {et~al.}(2017)\citenamefont {Flick},
  \citenamefont {Ruggenthaler}, \citenamefont {Appel},\ and\ \citenamefont
  {Rubio}}]{flick2017}%
  \BibitemOpen
  \bibfield  {author} {\bibinfo {author} {\bibfnamefont {J.}~\bibnamefont
  {Flick}}, \bibinfo {author} {\bibfnamefont {M.}~\bibnamefont {Ruggenthaler}},
  \bibinfo {author} {\bibfnamefont {H.}~\bibnamefont {Appel}}, \ and\ \bibinfo
  {author} {\bibfnamefont {A.}~\bibnamefont {Rubio}},\ }\bibfield  {title}
  {\enquote {\bibinfo {title} {Atoms and molecules in cavities, from weak to
  strong coupling in quantum-electrodynamics (qed) chemistry},}\ }\href@noop {}
  {\bibfield  {journal} {\bibinfo  {journal} {Proc. Natl. Acad. Sci. U.S.A.}\
  }\textbf {\bibinfo {volume} {114}},\ \bibinfo {pages} {3026--3034} (\bibinfo
  {year} {2017})}\BibitemShut {NoStop}%
\bibitem [{\citenamefont {Schäfer}\ \emph {et~al.}(2020)\citenamefont
  {Schäfer}, \citenamefont {Ruggenthaler}, \citenamefont {Rokaj},\ and\
  \citenamefont {Rubio}}]{schafer2020relevance}%
  \BibitemOpen
  \bibfield  {author} {\bibinfo {author} {\bibfnamefont {C.}~\bibnamefont
  {Schäfer}}, \bibinfo {author} {\bibfnamefont {M.}~\bibnamefont
  {Ruggenthaler}}, \bibinfo {author} {\bibfnamefont {V.}~\bibnamefont {Rokaj}},
  \ and\ \bibinfo {author} {\bibfnamefont {A.}~\bibnamefont {Rubio}},\
  }\bibfield  {title} {\enquote {\bibinfo {title} {Relevance of the quadratic
  diamagnetic and self-polarization terms in cavity quantum electrodynamics},}\
  }\href@noop {} {\bibfield  {journal} {\bibinfo  {journal} {ACS Photonics}\
  }\textbf {\bibinfo {volume} {7}},\ \bibinfo {pages} {975--990} (\bibinfo
  {year} {2020})}\BibitemShut {NoStop}%
\bibitem [{\citenamefont {Flick}\ \emph {et~al.}(2018)\citenamefont {Flick},
  \citenamefont {Schäfer}, \citenamefont {Ruggenthaler}, \citenamefont
  {Appel},\ and\ \citenamefont {Rubio}}]{flick2018abinito}%
  \BibitemOpen
  \bibfield  {author} {\bibinfo {author} {\bibfnamefont {J.}~\bibnamefont
  {Flick}}, \bibinfo {author} {\bibfnamefont {C.}~\bibnamefont {Schäfer}},
  \bibinfo {author} {\bibfnamefont {M.}~\bibnamefont {Ruggenthaler}}, \bibinfo
  {author} {\bibfnamefont {H.}~\bibnamefont {Appel}}, \ and\ \bibinfo {author}
  {\bibfnamefont {A.}~\bibnamefont {Rubio}},\ }\bibfield  {title} {\enquote
  {\bibinfo {title} {Ab initio optimized effective potentials for real
  molecules in optical cavities: Photon contributions to the molecular ground
  state},}\ }\href@noop {} {\bibfield  {journal} {\bibinfo  {journal} {ACS
  Photonics}\ }\textbf {\bibinfo {volume} {5}},\ \bibinfo {pages} {992--1005}
  (\bibinfo {year} {2018})}\BibitemShut {NoStop}%
\bibitem [{\citenamefont {Cohen}, \citenamefont {Mori-S{\'a}nchez},\ and\
  \citenamefont {Yang}(2008)}]{cohen2008insights}%
  \BibitemOpen
  \bibfield  {author} {\bibinfo {author} {\bibfnamefont {A.~J.}\ \bibnamefont
  {Cohen}}, \bibinfo {author} {\bibfnamefont {P.}~\bibnamefont
  {Mori-S{\'a}nchez}}, \ and\ \bibinfo {author} {\bibfnamefont
  {W.}~\bibnamefont {Yang}},\ }\bibfield  {title} {\enquote {\bibinfo {title}
  {Insights into current limitations of density functional theory},}\
  }\href@noop {} {\bibfield  {journal} {\bibinfo  {journal} {Science}\ }\textbf
  {\bibinfo {volume} {321}},\ \bibinfo {pages} {792--794} (\bibinfo {year}
  {2008})}\BibitemShut {NoStop}%
\bibitem [{\citenamefont {Hermann}, \citenamefont {DiStasio},\ and\
  \citenamefont {Tkatchenko}(2017)}]{hermann2017}%
  \BibitemOpen
  \bibfield  {author} {\bibinfo {author} {\bibfnamefont {J.}~\bibnamefont
  {Hermann}}, \bibinfo {author} {\bibfnamefont {R.~A.}\ \bibnamefont
  {DiStasio}}, \ and\ \bibinfo {author} {\bibfnamefont {A.}~\bibnamefont
  {Tkatchenko}},\ }\bibfield  {title} {\enquote {\bibinfo {title}
  {First-principles models for van der waals interactions in molecules and
  materials: Concepts, theory, and applications},}\ }\href@noop {} {\bibfield
  {journal} {\bibinfo  {journal} {Chem. Rev.}\ }\textbf {\bibinfo {volume}
  {117}},\ \bibinfo {pages} {4714--4758} (\bibinfo {year} {2017})}\BibitemShut
  {NoStop}%
\bibitem [{\citenamefont {Pellegrini}\ \emph {et~al.}(2015)\citenamefont
  {Pellegrini}, \citenamefont {Flick}, \citenamefont {Tokatly}, \citenamefont
  {Appel},\ and\ \citenamefont {Rubio}}]{pellegrini2015}%
  \BibitemOpen
  \bibfield  {author} {\bibinfo {author} {\bibfnamefont {C.}~\bibnamefont
  {Pellegrini}}, \bibinfo {author} {\bibfnamefont {J.}~\bibnamefont {Flick}},
  \bibinfo {author} {\bibfnamefont {I.~V.}\ \bibnamefont {Tokatly}}, \bibinfo
  {author} {\bibfnamefont {H.}~\bibnamefont {Appel}}, \ and\ \bibinfo {author}
  {\bibfnamefont {A.}~\bibnamefont {Rubio}},\ }\bibfield  {title} {\enquote
  {\bibinfo {title} {Optimized effective potential for quantum electrodynamical
  time-dependent density functional theory},}\ }\href@noop {} {\bibfield
  {journal} {\bibinfo  {journal} {Phys. Rev. Lett.}\ }\textbf {\bibinfo
  {volume} {115}},\ \bibinfo {pages} {093001} (\bibinfo {year}
  {2015})}\BibitemShut {NoStop}%
\bibitem [{\citenamefont {Sch{\"a}fer}\ \emph
  {et~al.}(2021{\natexlab{b}})\citenamefont {Sch{\"a}fer}, \citenamefont
  {Buchholz}, \citenamefont {Penz}, \citenamefont {Ruggenthaler},\ and\
  \citenamefont {Rubio}}]{Schafere2110464118}%
  \BibitemOpen
  \bibfield  {author} {\bibinfo {author} {\bibfnamefont {C.}~\bibnamefont
  {Sch{\"a}fer}}, \bibinfo {author} {\bibfnamefont {F.}~\bibnamefont
  {Buchholz}}, \bibinfo {author} {\bibfnamefont {M.}~\bibnamefont {Penz}},
  \bibinfo {author} {\bibfnamefont {M.}~\bibnamefont {Ruggenthaler}}, \ and\
  \bibinfo {author} {\bibfnamefont {A.}~\bibnamefont {Rubio}},\ }\bibfield
  {title} {\enquote {\bibinfo {title} {Making ab initio qed functional (s):
  Non-perturbative and photon-free effective frameworks for strong light-matter
  coupling},}\ }\href {\doibase 10.1073/pnas.2110464118} {\bibfield  {journal}
  {\bibinfo  {journal} {Proc. Natl. Acad. Sci. U.S.A.}\ }\textbf {\bibinfo
  {volume} {118}} (\bibinfo {year} {2021}{\natexlab{b}}),\
  10.1073/pnas.2110464118}\BibitemShut {NoStop}%
\bibitem [{\citenamefont {Mordovina}\ \emph {et~al.}(2020)\citenamefont
  {Mordovina}, \citenamefont {Bungey}, \citenamefont {Appel}, \citenamefont
  {Knowles}, \citenamefont {Rubio},\ and\ \citenamefont
  {Manby}}]{mordovina2020polaritonic}%
  \BibitemOpen
  \bibfield  {author} {\bibinfo {author} {\bibfnamefont {U.}~\bibnamefont
  {Mordovina}}, \bibinfo {author} {\bibfnamefont {C.}~\bibnamefont {Bungey}},
  \bibinfo {author} {\bibfnamefont {H.}~\bibnamefont {Appel}}, \bibinfo
  {author} {\bibfnamefont {P.~J.}\ \bibnamefont {Knowles}}, \bibinfo {author}
  {\bibfnamefont {A.}~\bibnamefont {Rubio}}, \ and\ \bibinfo {author}
  {\bibfnamefont {F.~R.}\ \bibnamefont {Manby}},\ }\bibfield  {title} {\enquote
  {\bibinfo {title} {Polaritonic coupled-cluster theory},}\ }\href@noop {}
  {\bibfield  {journal} {\bibinfo  {journal} {Phys. Rev. Res.}\ }\textbf
  {\bibinfo {volume} {2}},\ \bibinfo {pages} {023262} (\bibinfo {year}
  {2020})}\BibitemShut {NoStop}%
\bibitem [{\citenamefont {Haugland}\ \emph {et~al.}(2020)\citenamefont
  {Haugland}, \citenamefont {Ronca}, \citenamefont {Kj{\o}nstad}, \citenamefont
  {Rubio},\ and\ \citenamefont {Koch}}]{haugland2020coupled}%
  \BibitemOpen
  \bibfield  {author} {\bibinfo {author} {\bibfnamefont {T.~S.}\ \bibnamefont
  {Haugland}}, \bibinfo {author} {\bibfnamefont {E.}~\bibnamefont {Ronca}},
  \bibinfo {author} {\bibfnamefont {E.~F.}\ \bibnamefont {Kj{\o}nstad}},
  \bibinfo {author} {\bibfnamefont {A.}~\bibnamefont {Rubio}}, \ and\ \bibinfo
  {author} {\bibfnamefont {H.}~\bibnamefont {Koch}},\ }\bibfield  {title}
  {\enquote {\bibinfo {title} {Coupled cluster theory for molecular polaritons:
  Changing ground and excited states},}\ }\href@noop {} {\bibfield  {journal}
  {\bibinfo  {journal} {Phys. Rev. X}\ }\textbf {\bibinfo {volume} {10}},\
  \bibinfo {pages} {041043} (\bibinfo {year} {2020})}\BibitemShut {NoStop}%
\bibitem [{\citenamefont {Bartlett}\ and\ \citenamefont
  {Musia{\l}}(2007)}]{bartlett2007coupled}%
  \BibitemOpen
  \bibfield  {author} {\bibinfo {author} {\bibfnamefont {R.~J.}\ \bibnamefont
  {Bartlett}}\ and\ \bibinfo {author} {\bibfnamefont {M.}~\bibnamefont
  {Musia{\l}}},\ }\bibfield  {title} {\enquote {\bibinfo {title}
  {Coupled-cluster theory in quantum chemistry},}\ }\href@noop {} {\bibfield
  {journal} {\bibinfo  {journal} {Rev. Mod. Phys.}\ }\textbf {\bibinfo {volume}
  {79}},\ \bibinfo {pages} {291} (\bibinfo {year} {2007})}\BibitemShut
  {NoStop}%
\bibitem [{\citenamefont {Haugland}\ \emph {et~al.}(2021)\citenamefont
  {Haugland}, \citenamefont {Sch{\"a}fer}, \citenamefont {Ronca}, \citenamefont
  {Rubio},\ and\ \citenamefont {Koch}}]{haugland2021intermolecular}%
  \BibitemOpen
  \bibfield  {author} {\bibinfo {author} {\bibfnamefont {T.~S.}\ \bibnamefont
  {Haugland}}, \bibinfo {author} {\bibfnamefont {C.}~\bibnamefont
  {Sch{\"a}fer}}, \bibinfo {author} {\bibfnamefont {E.}~\bibnamefont {Ronca}},
  \bibinfo {author} {\bibfnamefont {A.}~\bibnamefont {Rubio}}, \ and\ \bibinfo
  {author} {\bibfnamefont {H.}~\bibnamefont {Koch}},\ }\bibfield  {title}
  {\enquote {\bibinfo {title} {Intermolecular interactions in optical cavities:
  An ab initio qed study},}\ }\href@noop {} {\bibfield  {journal} {\bibinfo
  {journal} {J. Chem. Phys.}\ }\textbf {\bibinfo {volume} {154}},\ \bibinfo
  {pages} {094113} (\bibinfo {year} {2021})}\BibitemShut {NoStop}%
\bibitem [{\citenamefont {DePrince~III}(2021)}]{deprince2021cavity}%
  \BibitemOpen
  \bibfield  {author} {\bibinfo {author} {\bibfnamefont {A.~E.}\ \bibnamefont
  {DePrince~III}},\ }\bibfield  {title} {\enquote {\bibinfo {title}
  {Cavity-modulated ionization potentials and electron affinities from quantum
  electrodynamics coupled-cluster theory},}\ }\href@noop {} {\bibfield
  {journal} {\bibinfo  {journal} {J. Chem. Phys.}\ }\textbf {\bibinfo {volume}
  {154}},\ \bibinfo {pages} {094112} (\bibinfo {year} {2021})}\BibitemShut
  {NoStop}%
\bibitem [{\citenamefont {Pavošević}\ and\ \citenamefont
  {Flick}(2021)}]{Pavosevic2021}%
  \BibitemOpen
  \bibfield  {author} {\bibinfo {author} {\bibfnamefont {F.}~\bibnamefont
  {Pavošević}}\ and\ \bibinfo {author} {\bibfnamefont {J.}~\bibnamefont
  {Flick}},\ }\bibfield  {title} {\enquote {\bibinfo {title} {Polaritonic
  unitary coupled cluster for quantum computations},}\ }\href {\doibase
  10.1021/acs.jpclett.1c02659} {\bibfield  {journal} {\bibinfo  {journal} {J.
  Phys. Chem. Lett.}\ }\textbf {\bibinfo {volume} {12}},\ \bibinfo {pages}
  {9100--9107} (\bibinfo {year} {2021})}\BibitemShut {NoStop}%
\bibitem [{\citenamefont {Liebenthal}, \citenamefont {Vu},\ and\ \citenamefont
  {DePrince}(2022)}]{liebenthal2021equationofmotion}%
  \BibitemOpen
  \bibfield  {author} {\bibinfo {author} {\bibfnamefont {M.~D.}\ \bibnamefont
  {Liebenthal}}, \bibinfo {author} {\bibfnamefont {N.}~\bibnamefont {Vu}}, \
  and\ \bibinfo {author} {\bibfnamefont {A.~E.}\ \bibnamefont {DePrince}},\
  }\bibfield  {title} {\enquote {\bibinfo {title} {Equation-of-motion cavity
  quantum electrodynamics coupled-cluster theory for electron attachment},}\
  }\href@noop {} {\bibfield  {journal} {\bibinfo  {journal} {J. Chem. Phys.}\
  }\textbf {\bibinfo {volume} {156}},\ \bibinfo {pages} {054105} (\bibinfo
  {year} {2022})}\BibitemShut {NoStop}%
\bibitem [{\citenamefont {Pavo{\v{s}}evi{\'c}}\ \emph
  {et~al.}(2022)\citenamefont {Pavo{\v{s}}evi{\'c}}, \citenamefont
  {Hammes-Schiffer}, \citenamefont {Rubio},\ and\ \citenamefont
  {Flick}}]{pavosevic2021cavity}%
  \BibitemOpen
  \bibfield  {author} {\bibinfo {author} {\bibfnamefont {F.}~\bibnamefont
  {Pavo{\v{s}}evi{\'c}}}, \bibinfo {author} {\bibfnamefont {S.}~\bibnamefont
  {Hammes-Schiffer}}, \bibinfo {author} {\bibfnamefont {A.}~\bibnamefont
  {Rubio}}, \ and\ \bibinfo {author} {\bibfnamefont {J.}~\bibnamefont
  {Flick}},\ }\bibfield  {title} {\enquote {\bibinfo {title} {Cavity-modulated
  proton transfer reactions},}\ }\href@noop {} {\bibfield  {journal} {\bibinfo
  {journal} {J. Am. Chem. Soc.}\ }\textbf {\bibinfo {volume} {144}},\ \bibinfo
  {pages} {4995–5002} (\bibinfo {year} {2022})}\BibitemShut {NoStop}%
\bibitem [{\citenamefont {Jones}\ \emph {et~al.}(2020)\citenamefont {Jones},
  \citenamefont {Mosquera}, \citenamefont {Schatz},\ and\ \citenamefont
  {Ratner}}]{jones2020embedding}%
  \BibitemOpen
  \bibfield  {author} {\bibinfo {author} {\bibfnamefont {L.~O.}\ \bibnamefont
  {Jones}}, \bibinfo {author} {\bibfnamefont {M.~A.}\ \bibnamefont {Mosquera}},
  \bibinfo {author} {\bibfnamefont {G.~C.}\ \bibnamefont {Schatz}}, \ and\
  \bibinfo {author} {\bibfnamefont {M.~A.}\ \bibnamefont {Ratner}},\ }\bibfield
   {title} {\enquote {\bibinfo {title} {Embedding methods for quantum
  chemistry: Applications from materials to life sciences},}\ }\href@noop {}
  {\bibfield  {journal} {\bibinfo  {journal} {J. Am. Chem. Soc.}\ }\textbf
  {\bibinfo {volume} {142}},\ \bibinfo {pages} {3281--3295} (\bibinfo {year}
  {2020})}\BibitemShut {NoStop}%
\bibitem [{\citenamefont {Sun}\ and\ \citenamefont
  {Chan}(2016)}]{sun2016quantum}%
  \BibitemOpen
  \bibfield  {author} {\bibinfo {author} {\bibfnamefont {Q.}~\bibnamefont
  {Sun}}\ and\ \bibinfo {author} {\bibfnamefont {G.~K.-L.}\ \bibnamefont
  {Chan}},\ }\bibfield  {title} {\enquote {\bibinfo {title} {Quantum embedding
  theories},}\ }\href@noop {} {\bibfield  {journal} {\bibinfo  {journal} {Acc.
  Chem. Res.}\ }\textbf {\bibinfo {volume} {49}},\ \bibinfo {pages}
  {2705--2712} (\bibinfo {year} {2016})}\BibitemShut {NoStop}%
\bibitem [{\citenamefont {Reinhard}\ \emph {et~al.}(2019)\citenamefont
  {Reinhard}, \citenamefont {Mordovina}, \citenamefont {Hubig}, \citenamefont
  {Kretchmer}, \citenamefont {Schollwöck}, \citenamefont {Appel},
  \citenamefont {Sentef},\ and\ \citenamefont {Rubio}}]{reinhard2019density}%
  \BibitemOpen
  \bibfield  {author} {\bibinfo {author} {\bibfnamefont {T.~E.}\ \bibnamefont
  {Reinhard}}, \bibinfo {author} {\bibfnamefont {U.}~\bibnamefont {Mordovina}},
  \bibinfo {author} {\bibfnamefont {C.}~\bibnamefont {Hubig}}, \bibinfo
  {author} {\bibfnamefont {J.~S.}\ \bibnamefont {Kretchmer}}, \bibinfo {author}
  {\bibfnamefont {U.}~\bibnamefont {Schollwöck}}, \bibinfo {author}
  {\bibfnamefont {H.}~\bibnamefont {Appel}}, \bibinfo {author} {\bibfnamefont
  {M.~A.}\ \bibnamefont {Sentef}}, \ and\ \bibinfo {author} {\bibfnamefont
  {A.}~\bibnamefont {Rubio}},\ }\bibfield  {title} {\enquote {\bibinfo {title}
  {Density-matrix embedding theory study of the one-dimensional
  hubbard--holstein model},}\ }\href@noop {} {\bibfield  {journal} {\bibinfo
  {journal} {J. Chem. Theory Comput.}\ }\textbf {\bibinfo {volume} {15}},\
  \bibinfo {pages} {2221--2232} (\bibinfo {year} {2019})}\BibitemShut {NoStop}%
\bibitem [{\citenamefont {Manby}\ \emph {et~al.}(2012)\citenamefont {Manby},
  \citenamefont {Stella}, \citenamefont {Goodpaster},\ and\ \citenamefont
  {Miller~III}}]{manby2012simple}%
  \BibitemOpen
  \bibfield  {author} {\bibinfo {author} {\bibfnamefont {F.~R.}\ \bibnamefont
  {Manby}}, \bibinfo {author} {\bibfnamefont {M.}~\bibnamefont {Stella}},
  \bibinfo {author} {\bibfnamefont {J.~D.}\ \bibnamefont {Goodpaster}}, \ and\
  \bibinfo {author} {\bibfnamefont {T.~F.}\ \bibnamefont {Miller~III}},\
  }\bibfield  {title} {\enquote {\bibinfo {title} {A simple, exact
  density-functional-theory embedding scheme},}\ }\href@noop {} {\bibfield
  {journal} {\bibinfo  {journal} {J. Chem. Theory Comput.}\ }\textbf {\bibinfo
  {volume} {8}},\ \bibinfo {pages} {2564--2568} (\bibinfo {year}
  {2012})}\BibitemShut {NoStop}%
\bibitem [{\citenamefont {Lee}\ \emph {et~al.}(2019)\citenamefont {Lee},
  \citenamefont {Welborn}, \citenamefont {Manby},\ and\ \citenamefont
  {Miller~III}}]{lee2019projection}%
  \BibitemOpen
  \bibfield  {author} {\bibinfo {author} {\bibfnamefont {S.~J.}\ \bibnamefont
  {Lee}}, \bibinfo {author} {\bibfnamefont {M.}~\bibnamefont {Welborn}},
  \bibinfo {author} {\bibfnamefont {F.~R.}\ \bibnamefont {Manby}}, \ and\
  \bibinfo {author} {\bibfnamefont {T.~F.}\ \bibnamefont {Miller~III}},\
  }\bibfield  {title} {\enquote {\bibinfo {title} {Projection-based
  wavefunction-in-dft embedding},}\ }\href@noop {} {\bibfield  {journal}
  {\bibinfo  {journal} {Acc. Chem. Res.}\ }\textbf {\bibinfo {volume} {52}},\
  \bibinfo {pages} {1359--1368} (\bibinfo {year} {2019})}\BibitemShut {NoStop}%
\bibitem [{\citenamefont {Bennie}\ \emph {et~al.}(2015)\citenamefont {Bennie},
  \citenamefont {Stella}, \citenamefont {Miller~III},\ and\ \citenamefont
  {Manby}}]{bennie2015accelerating}%
  \BibitemOpen
  \bibfield  {author} {\bibinfo {author} {\bibfnamefont {S.~J.}\ \bibnamefont
  {Bennie}}, \bibinfo {author} {\bibfnamefont {M.}~\bibnamefont {Stella}},
  \bibinfo {author} {\bibfnamefont {T.~F.}\ \bibnamefont {Miller~III}}, \ and\
  \bibinfo {author} {\bibfnamefont {F.~R.}\ \bibnamefont {Manby}},\ }\bibfield
  {title} {\enquote {\bibinfo {title} {Accelerating wavefunction in
  density-functional-theory embedding by truncating the active basis set},}\
  }\href@noop {} {\bibfield  {journal} {\bibinfo  {journal} {J. Chem. Phys.}\
  }\textbf {\bibinfo {volume} {143}},\ \bibinfo {pages} {024105} (\bibinfo
  {year} {2015})}\BibitemShut {NoStop}%
\bibitem [{\citenamefont {Bensberg}\ and\ \citenamefont
  {Neugebauer}(2019)}]{bensberg2019automatic}%
  \BibitemOpen
  \bibfield  {author} {\bibinfo {author} {\bibfnamefont {M.}~\bibnamefont
  {Bensberg}}\ and\ \bibinfo {author} {\bibfnamefont {J.}~\bibnamefont
  {Neugebauer}},\ }\bibfield  {title} {\enquote {\bibinfo {title} {Automatic
  basis-set adaptation in projection-based embedding},}\ }\href@noop {}
  {\bibfield  {journal} {\bibinfo  {journal} {J. Chem. Phys.}\ }\textbf
  {\bibinfo {volume} {150}},\ \bibinfo {pages} {184104} (\bibinfo {year}
  {2019})}\BibitemShut {NoStop}%
\bibitem [{\citenamefont {Claudino}\ and\ \citenamefont
  {Mayhall}(2019{\natexlab{a}})}]{claudino2019simple}%
  \BibitemOpen
  \bibfield  {author} {\bibinfo {author} {\bibfnamefont {D.}~\bibnamefont
  {Claudino}}\ and\ \bibinfo {author} {\bibfnamefont {N.~J.}\ \bibnamefont
  {Mayhall}},\ }\bibfield  {title} {\enquote {\bibinfo {title} {Simple and
  efficient truncation of virtual spaces in embedded wave functions via
  concentric localization},}\ }\href@noop {} {\bibfield  {journal} {\bibinfo
  {journal} {J. Chem. Theory Comput.}\ }\textbf {\bibinfo {volume} {15}},\
  \bibinfo {pages} {6085--6096} (\bibinfo {year}
  {2019}{\natexlab{a}})}\BibitemShut {NoStop}%
\bibitem [{\citenamefont {Werner}\ \emph {et~al.}(2020)\citenamefont {Werner},
  \citenamefont {Knowles}, \citenamefont {Manby}, \citenamefont {Black},
  \citenamefont {Doll}, \citenamefont {He{\ss}elmann}, \citenamefont {Kats},
  \citenamefont {K{\"o}hn}, \citenamefont {Korona}, \citenamefont {Kreplin}
  \emph {et~al.}}]{werner2020molpro}%
  \BibitemOpen
  \bibfield  {author} {\bibinfo {author} {\bibfnamefont {H.-J.}\ \bibnamefont
  {Werner}}, \bibinfo {author} {\bibfnamefont {P.~J.}\ \bibnamefont {Knowles}},
  \bibinfo {author} {\bibfnamefont {F.~R.}\ \bibnamefont {Manby}}, \bibinfo
  {author} {\bibfnamefont {J.~A.}\ \bibnamefont {Black}}, \bibinfo {author}
  {\bibfnamefont {K.}~\bibnamefont {Doll}}, \bibinfo {author} {\bibfnamefont
  {A.}~\bibnamefont {He{\ss}elmann}}, \bibinfo {author} {\bibfnamefont
  {D.}~\bibnamefont {Kats}}, \bibinfo {author} {\bibfnamefont {A.}~\bibnamefont
  {K{\"o}hn}}, \bibinfo {author} {\bibfnamefont {T.}~\bibnamefont {Korona}},
  \bibinfo {author} {\bibfnamefont {D.~A.}\ \bibnamefont {Kreplin}},  \emph
  {et~al.},\ }\bibfield  {title} {\enquote {\bibinfo {title} {The molpro
  quantum chemistry package},}\ }\href@noop {} {\bibfield  {journal} {\bibinfo
  {journal} {J. Chem. Phys.}\ }\textbf {\bibinfo {volume} {152}},\ \bibinfo
  {pages} {144107} (\bibinfo {year} {2020})}\BibitemShut {NoStop}%
\bibitem [{\citenamefont {Craig}\ and\ \citenamefont
  {Thirunamachandran}(1998)}]{craig1998}%
  \BibitemOpen
  \bibfield  {author} {\bibinfo {author} {\bibfnamefont {D.}~\bibnamefont
  {Craig}}\ and\ \bibinfo {author} {\bibfnamefont {T.}~\bibnamefont
  {Thirunamachandran}},\ }\href@noop {} {\emph {\bibinfo {title} {Molecular
  Quantum Electrodynamics: An Introduction to Radiation-molecule
  Interactions}}}\ (\bibinfo  {publisher} {Dover Publications},\ \bibinfo
  {year} {1998})\BibitemShut {NoStop}%
\bibitem [{\citenamefont {Pavo\v{s}evi\'{c}}, \citenamefont {Culpitt},\ and\
  \citenamefont {Hammes-Schiffer}(2019)}]{Hammes-Schiffer19_338}%
  \BibitemOpen
  \bibfield  {author} {\bibinfo {author} {\bibfnamefont {F.}~\bibnamefont
  {Pavo\v{s}evi\'{c}}}, \bibinfo {author} {\bibfnamefont {T.}~\bibnamefont
  {Culpitt}}, \ and\ \bibinfo {author} {\bibfnamefont {S.}~\bibnamefont
  {Hammes-Schiffer}},\ }\bibfield  {title} {\enquote {\bibinfo {title}
  {Multicomponent coupled cluster singles and doubles theory within the
  nuclear-electronic orbital framework},}\ }\href@noop {} {\bibfield  {journal}
  {\bibinfo  {journal} {J. Chem. Theory Comput.}\ }\textbf {\bibinfo {volume}
  {15}},\ \bibinfo {pages} {338--347} (\bibinfo {year} {2019})}\BibitemShut
  {NoStop}%
\bibitem [{\citenamefont {Pavo\v{s}evi\'{c}}, \citenamefont {Culpitt},\ and\
  \citenamefont {Hammes-Schiffer}(2020)}]{Hammes-Schiffer20_4222}%
  \BibitemOpen
  \bibfield  {author} {\bibinfo {author} {\bibfnamefont {F.}~\bibnamefont
  {Pavo\v{s}evi\'{c}}}, \bibinfo {author} {\bibfnamefont {T.}~\bibnamefont
  {Culpitt}}, \ and\ \bibinfo {author} {\bibfnamefont {S.}~\bibnamefont
  {Hammes-Schiffer}},\ }\bibfield  {title} {\enquote {\bibinfo {title}
  {Multicomponent quantum chemistry: Integrating electronic and nuclear quantum
  effects via the nuclear-electronic orbital method},}\ }\href@noop {}
  {\bibfield  {journal} {\bibinfo  {journal} {Chem. Rev.}\ }\textbf {\bibinfo
  {volume} {120}},\ \bibinfo {pages} {4222--4253} (\bibinfo {year}
  {2020})}\BibitemShut {NoStop}%
\bibitem [{\citenamefont {Pavo{\v{s}}evi{\'c}}\ and\ \citenamefont
  {Hammes-Schiffer}(2021)}]{pavosevic2021multicomponent}%
  \BibitemOpen
  \bibfield  {author} {\bibinfo {author} {\bibfnamefont {F.}~\bibnamefont
  {Pavo{\v{s}}evi{\'c}}}\ and\ \bibinfo {author} {\bibfnamefont
  {S.}~\bibnamefont {Hammes-Schiffer}},\ }\bibfield  {title} {\enquote
  {\bibinfo {title} {Multicomponent unitary coupled cluster and
  equation-of-motion for quantum computation.}}\ }\href@noop {} {\bibfield
  {journal} {\bibinfo  {journal} {J. Chem. Theory Comput.}\ }\textbf {\bibinfo
  {volume} {17}},\ \bibinfo {pages} {3252–3258} (\bibinfo {year}
  {2021})}\BibitemShut {NoStop}%
\bibitem [{\citenamefont {Galego}\ \emph {et~al.}(2019)\citenamefont {Galego},
  \citenamefont {Climent}, \citenamefont {Garcia-Vidal},\ and\ \citenamefont
  {Feist}}]{Galego2019}%
  \BibitemOpen
  \bibfield  {author} {\bibinfo {author} {\bibfnamefont {J.}~\bibnamefont
  {Galego}}, \bibinfo {author} {\bibfnamefont {C.}~\bibnamefont {Climent}},
  \bibinfo {author} {\bibfnamefont {F.~J.}\ \bibnamefont {Garcia-Vidal}}, \
  and\ \bibinfo {author} {\bibfnamefont {J.}~\bibnamefont {Feist}},\ }\bibfield
   {title} {\enquote {\bibinfo {title} {Cavity casimir-polder forces and their
  effects in ground-state chemical reactivity},}\ }\href@noop {} {\bibfield
  {journal} {\bibinfo  {journal} {Phys. Rev. X}\ }\textbf {\bibinfo {volume}
  {9}},\ \bibinfo {pages} {021057} (\bibinfo {year} {2019})}\BibitemShut
  {NoStop}%
\bibitem [{\citenamefont {Benz}\ \emph {et~al.}(2016)\citenamefont {Benz},
  \citenamefont {Schmidt}, \citenamefont {Dreismann}, \citenamefont
  {Chikkaraddy}, \citenamefont {Zhang}, \citenamefont {Demetriadou},
  \citenamefont {Carnegie}, \citenamefont {Ohadi}, \citenamefont {de~Nijs},
  \citenamefont {Esteban}, \citenamefont {Aizpurua},\ and\ \citenamefont
  {Baumberg}}]{benz20216}%
  \BibitemOpen
  \bibfield  {author} {\bibinfo {author} {\bibfnamefont {F.}~\bibnamefont
  {Benz}}, \bibinfo {author} {\bibfnamefont {M.~K.}\ \bibnamefont {Schmidt}},
  \bibinfo {author} {\bibfnamefont {A.}~\bibnamefont {Dreismann}}, \bibinfo
  {author} {\bibfnamefont {R.}~\bibnamefont {Chikkaraddy}}, \bibinfo {author}
  {\bibfnamefont {Y.}~\bibnamefont {Zhang}}, \bibinfo {author} {\bibfnamefont
  {A.}~\bibnamefont {Demetriadou}}, \bibinfo {author} {\bibfnamefont
  {C.}~\bibnamefont {Carnegie}}, \bibinfo {author} {\bibfnamefont
  {H.}~\bibnamefont {Ohadi}}, \bibinfo {author} {\bibfnamefont
  {B.}~\bibnamefont {de~Nijs}}, \bibinfo {author} {\bibfnamefont
  {R.}~\bibnamefont {Esteban}}, \bibinfo {author} {\bibfnamefont
  {J.}~\bibnamefont {Aizpurua}}, \ and\ \bibinfo {author} {\bibfnamefont
  {J.~J.}\ \bibnamefont {Baumberg}},\ }\bibfield  {title} {\enquote {\bibinfo
  {title} {Single-molecule optomechanics in picocavities},}\ }\href@noop {}
  {\bibfield  {journal} {\bibinfo  {journal} {Science}\ }\textbf {\bibinfo
  {volume} {354}},\ \bibinfo {pages} {726--729} (\bibinfo {year}
  {2016})}\BibitemShut {NoStop}%
\bibitem [{\citenamefont {Richard-Lacroix}\ and\ \citenamefont
  {Deckert}(2020)}]{Richard_Lacroix_2020}%
  \BibitemOpen
  \bibfield  {author} {\bibinfo {author} {\bibfnamefont {M.}~\bibnamefont
  {Richard-Lacroix}}\ and\ \bibinfo {author} {\bibfnamefont {V.}~\bibnamefont
  {Deckert}},\ }\bibfield  {title} {\enquote {\bibinfo {title} {Direct
  molecular-level near-field plasmon and temperature assessment in a single
  plasmonic hotspot},}\ }\href@noop {} {\bibfield  {journal} {\bibinfo
  {journal} {Light: Science {\&} Applications}\ }\textbf {\bibinfo {volume}
  {9}} (\bibinfo {year} {2020})}\BibitemShut {NoStop}%
\bibitem [{\citenamefont {Liu}\ \emph {et~al.}(2021)\citenamefont {Liu},
  \citenamefont {Hammud}, \citenamefont {Wolf},\ and\ \citenamefont
  {Kumagai}}]{Liu2021}%
  \BibitemOpen
  \bibfield  {author} {\bibinfo {author} {\bibfnamefont {S.}~\bibnamefont
  {Liu}}, \bibinfo {author} {\bibfnamefont {A.}~\bibnamefont {Hammud}},
  \bibinfo {author} {\bibfnamefont {M.}~\bibnamefont {Wolf}}, \ and\ \bibinfo
  {author} {\bibfnamefont {T.}~\bibnamefont {Kumagai}},\ }\bibfield  {title}
  {\enquote {\bibinfo {title} {Atomic point contact raman spectroscopy of a
  si(111)-7 × 7 surface},}\ }\href@noop {} {\bibfield  {journal} {\bibinfo
  {journal} {Nano Lett.}\ }\textbf {\bibinfo {volume} {21}},\ \bibinfo {pages}
  {4057--4061} (\bibinfo {year} {2021})}\BibitemShut {NoStop}%
\bibitem [{\citenamefont {Griffiths}\ \emph {et~al.}(2021)\citenamefont
  {Griffiths}, \citenamefont {de~Nijs}, \citenamefont {Chikkaraddy},\ and\
  \citenamefont {Baumberg}}]{griffiths2021}%
  \BibitemOpen
  \bibfield  {author} {\bibinfo {author} {\bibfnamefont {J.}~\bibnamefont
  {Griffiths}}, \bibinfo {author} {\bibfnamefont {B.}~\bibnamefont {de~Nijs}},
  \bibinfo {author} {\bibfnamefont {R.}~\bibnamefont {Chikkaraddy}}, \ and\
  \bibinfo {author} {\bibfnamefont {J.~J.}\ \bibnamefont {Baumberg}},\
  }\bibfield  {title} {\enquote {\bibinfo {title} {Locating single-atom optical
  picocavities using wavelength-multiplexed raman scattering},}\ }\href@noop {}
  {\bibfield  {journal} {\bibinfo  {journal} {ACS Photonics}\ }\textbf
  {\bibinfo {volume} {8}},\ \bibinfo {pages} {2868--2875} (\bibinfo {year}
  {2021})}\BibitemShut {NoStop}%
\bibitem [{\citenamefont {Rokaj}\ \emph {et~al.}(2018)\citenamefont {Rokaj},
  \citenamefont {Welakuh}, \citenamefont {Ruggenthaler},\ and\ \citenamefont
  {Rubio}}]{rokaj2018}%
  \BibitemOpen
  \bibfield  {author} {\bibinfo {author} {\bibfnamefont {V.}~\bibnamefont
  {Rokaj}}, \bibinfo {author} {\bibfnamefont {D.~M.}\ \bibnamefont {Welakuh}},
  \bibinfo {author} {\bibfnamefont {M.}~\bibnamefont {Ruggenthaler}}, \ and\
  \bibinfo {author} {\bibfnamefont {A.}~\bibnamefont {Rubio}},\ }\bibfield
  {title} {\enquote {\bibinfo {title} {Light--matter interaction in the
  long-wavelength limit: No ground-state without dipole self-energy},}\
  }\href@noop {} {\bibfield  {journal} {\bibinfo  {journal} {J. Phys. B}\
  }\textbf {\bibinfo {volume} {51}},\ \bibinfo {pages} {034005} (\bibinfo
  {year} {2018})}\BibitemShut {NoStop}%
\bibitem [{\citenamefont {Taylor}\ \emph {et~al.}(2020)\citenamefont {Taylor},
  \citenamefont {Mandal}, \citenamefont {Zhou},\ and\ \citenamefont
  {Huo}}]{Taylor2020}%
  \BibitemOpen
  \bibfield  {author} {\bibinfo {author} {\bibfnamefont {M.~A.~D.}\
  \bibnamefont {Taylor}}, \bibinfo {author} {\bibfnamefont {A.}~\bibnamefont
  {Mandal}}, \bibinfo {author} {\bibfnamefont {W.}~\bibnamefont {Zhou}}, \ and\
  \bibinfo {author} {\bibfnamefont {P.}~\bibnamefont {Huo}},\ }\bibfield
  {title} {\enquote {\bibinfo {title} {Resolution of gauge ambiguities in
  molecular cavity quantum electrodynamics},}\ }\href@noop {} {\bibfield
  {journal} {\bibinfo  {journal} {Phys. Rev. Lett.}\ }\textbf {\bibinfo
  {volume} {125}},\ \bibinfo {pages} {123602} (\bibinfo {year}
  {2020})}\BibitemShut {NoStop}%
\bibitem [{\citenamefont {Andolina}\ \emph {et~al.}(2020)\citenamefont
  {Andolina}, \citenamefont {Pellegrino}, \citenamefont {Giovannetti},
  \citenamefont {MacDonald},\ and\ \citenamefont
  {Polini}}]{andolina2020theory}%
  \BibitemOpen
  \bibfield  {author} {\bibinfo {author} {\bibfnamefont {G.}~\bibnamefont
  {Andolina}}, \bibinfo {author} {\bibfnamefont {F.}~\bibnamefont
  {Pellegrino}}, \bibinfo {author} {\bibfnamefont {V.}~\bibnamefont
  {Giovannetti}}, \bibinfo {author} {\bibfnamefont {A.}~\bibnamefont
  {MacDonald}}, \ and\ \bibinfo {author} {\bibfnamefont {M.}~\bibnamefont
  {Polini}},\ }\bibfield  {title} {\enquote {\bibinfo {title} {Theory of photon
  condensation in a spatially varying electromagnetic field},}\ }\href@noop {}
  {\bibfield  {journal} {\bibinfo  {journal} {Phys. Rev. B}\ }\textbf {\bibinfo
  {volume} {102}},\ \bibinfo {pages} {125137} (\bibinfo {year}
  {2020})}\BibitemShut {NoStop}%
\bibitem [{\citenamefont {Smith}\ \emph {et~al.}(2018)\citenamefont {Smith},
  \citenamefont {Burns}, \citenamefont {Sirianni}, \citenamefont {Nascimento},
  \citenamefont {Kumar}, \citenamefont {James}, \citenamefont {Schriber},
  \citenamefont {Zhang}, \citenamefont {Zhang}, \citenamefont {Abbott} \emph
  {et~al.}}]{smith2018psi4numpy}%
  \BibitemOpen
  \bibfield  {author} {\bibinfo {author} {\bibfnamefont {D.~G.}\ \bibnamefont
  {Smith}}, \bibinfo {author} {\bibfnamefont {L.~A.}\ \bibnamefont {Burns}},
  \bibinfo {author} {\bibfnamefont {D.~A.}\ \bibnamefont {Sirianni}}, \bibinfo
  {author} {\bibfnamefont {D.~R.}\ \bibnamefont {Nascimento}}, \bibinfo
  {author} {\bibfnamefont {A.}~\bibnamefont {Kumar}}, \bibinfo {author}
  {\bibfnamefont {A.~M.}\ \bibnamefont {James}}, \bibinfo {author}
  {\bibfnamefont {J.~B.}\ \bibnamefont {Schriber}}, \bibinfo {author}
  {\bibfnamefont {T.}~\bibnamefont {Zhang}}, \bibinfo {author} {\bibfnamefont
  {B.}~\bibnamefont {Zhang}}, \bibinfo {author} {\bibfnamefont {A.~S.}\
  \bibnamefont {Abbott}},  \emph {et~al.},\ }\bibfield  {title} {\enquote
  {\bibinfo {title} {Psi4numpy: An interactive quantum chemistry programming
  environment for reference implementations and rapid development},}\
  }\href@noop {} {\bibfield  {journal} {\bibinfo  {journal} {J. Chem. Theory
  Comput.}\ }\textbf {\bibinfo {volume} {14}},\ \bibinfo {pages} {3504--3511}
  (\bibinfo {year} {2018})}\BibitemShut {NoStop}%
\bibitem [{\citenamefont {Smith}\ \emph {et~al.}(2020)\citenamefont {Smith},
  \citenamefont {Burns}, \citenamefont {Sirianni}, \citenamefont {Nascimento},
  \citenamefont {Kumar}, \citenamefont {James}, \citenamefont {Schriber},
  \citenamefont {Zhang}, \citenamefont {Zhang}, \citenamefont {Abbott} \emph
  {et~al.}}]{githubpsi4numpy}%
  \BibitemOpen
  \bibfield  {author} {\bibinfo {author} {\bibfnamefont {D.~G.}\ \bibnamefont
  {Smith}}, \bibinfo {author} {\bibfnamefont {L.~A.}\ \bibnamefont {Burns}},
  \bibinfo {author} {\bibfnamefont {D.~A.}\ \bibnamefont {Sirianni}}, \bibinfo
  {author} {\bibfnamefont {D.~R.}\ \bibnamefont {Nascimento}}, \bibinfo
  {author} {\bibfnamefont {A.}~\bibnamefont {Kumar}}, \bibinfo {author}
  {\bibfnamefont {A.~M.}\ \bibnamefont {James}}, \bibinfo {author}
  {\bibfnamefont {J.~B.}\ \bibnamefont {Schriber}}, \bibinfo {author}
  {\bibfnamefont {T.}~\bibnamefont {Zhang}}, \bibinfo {author} {\bibfnamefont
  {B.}~\bibnamefont {Zhang}}, \bibinfo {author} {\bibfnamefont {A.~S.}\
  \bibnamefont {Abbott}},  \emph {et~al.},\ }\href@noop {} {\enquote {\bibinfo
  {title} {Psi4numpy: An interactive quantum chemistry programming environment
  for reference implementations and rapid development},}\ } (\bibinfo {year}
  {2020})\BibitemShut {NoStop}%
\bibitem [{\citenamefont {Ditchfield}, \citenamefont {Hehre},\ and\
  \citenamefont {Pople}(1971)}]{ditchfield1971self}%
  \BibitemOpen
  \bibfield  {author} {\bibinfo {author} {\bibfnamefont {R.}~\bibnamefont
  {Ditchfield}}, \bibinfo {author} {\bibfnamefont {W.~J.}\ \bibnamefont
  {Hehre}}, \ and\ \bibinfo {author} {\bibfnamefont {J.~A.}\ \bibnamefont
  {Pople}},\ }\bibfield  {title} {\enquote {\bibinfo {title} {Self-consistent
  molecular-orbital methods. ix. an extended gaussian-type basis for
  molecular-orbital studies of organic molecules},}\ }\href@noop {} {\bibfield
  {journal} {\bibinfo  {journal} {J. Chem. Phys.}\ }\textbf {\bibinfo {volume}
  {54}},\ \bibinfo {pages} {724--728} (\bibinfo {year} {1971})}\BibitemShut
  {NoStop}%
\bibitem [{\citenamefont {Hariharan}\ and\ \citenamefont
  {Pople}(1973)}]{hariharan1973influence}%
  \BibitemOpen
  \bibfield  {author} {\bibinfo {author} {\bibfnamefont {P.~C.}\ \bibnamefont
  {Hariharan}}\ and\ \bibinfo {author} {\bibfnamefont {J.~A.}\ \bibnamefont
  {Pople}},\ }\bibfield  {title} {\enquote {\bibinfo {title} {The influence of
  polarization functions on molecular orbital hydrogenation energies},}\
  }\href@noop {} {\bibfield  {journal} {\bibinfo  {journal} {Theor. Chim.
  Acta}\ }\textbf {\bibinfo {volume} {28}},\ \bibinfo {pages} {213--222}
  (\bibinfo {year} {1973})}\BibitemShut {NoStop}%
\bibitem [{\citenamefont {Rassolov}\ \emph {et~al.}(2001)\citenamefont
  {Rassolov}, \citenamefont {Ratner}, \citenamefont {Pople}, \citenamefont
  {Redfern},\ and\ \citenamefont {Curtiss}}]{rassolov20016}%
  \BibitemOpen
  \bibfield  {author} {\bibinfo {author} {\bibfnamefont {V.~A.}\ \bibnamefont
  {Rassolov}}, \bibinfo {author} {\bibfnamefont {M.~A.}\ \bibnamefont
  {Ratner}}, \bibinfo {author} {\bibfnamefont {J.~A.}\ \bibnamefont {Pople}},
  \bibinfo {author} {\bibfnamefont {P.~C.}\ \bibnamefont {Redfern}}, \ and\
  \bibinfo {author} {\bibfnamefont {L.~A.}\ \bibnamefont {Curtiss}},\
  }\bibfield  {title} {\enquote {\bibinfo {title} {6-31g* basis set for
  third-row atoms},}\ }\href@noop {} {\bibfield  {journal} {\bibinfo  {journal}
  {J. Comput. Chem.}\ }\textbf {\bibinfo {volume} {22}},\ \bibinfo {pages}
  {976--984} (\bibinfo {year} {2001})}\BibitemShut {NoStop}%
\bibitem [{\citenamefont {Neese}\ \emph {et~al.}(2020)\citenamefont {Neese},
  \citenamefont {Wennmohs}, \citenamefont {Becker},\ and\ \citenamefont
  {Riplinger}}]{neese2020orca}%
  \BibitemOpen
  \bibfield  {author} {\bibinfo {author} {\bibfnamefont {F.}~\bibnamefont
  {Neese}}, \bibinfo {author} {\bibfnamefont {F.}~\bibnamefont {Wennmohs}},
  \bibinfo {author} {\bibfnamefont {U.}~\bibnamefont {Becker}}, \ and\ \bibinfo
  {author} {\bibfnamefont {C.}~\bibnamefont {Riplinger}},\ }\bibfield  {title}
  {\enquote {\bibinfo {title} {The orca quantum chemistry program package},}\
  }\href@noop {} {\bibfield  {journal} {\bibinfo  {journal} {J. Chem. Phys.}\
  }\textbf {\bibinfo {volume} {152}},\ \bibinfo {pages} {224108} (\bibinfo
  {year} {2020})}\BibitemShut {NoStop}%
\bibitem [{\citenamefont {Dunning~Jr}(1989)}]{dunning1989gaussian}%
  \BibitemOpen
  \bibfield  {author} {\bibinfo {author} {\bibfnamefont {T.~H.}\ \bibnamefont
  {Dunning~Jr}},\ }\bibfield  {title} {\enquote {\bibinfo {title} {Gaussian
  basis sets for use in correlated molecular calculations. i. the atoms boron
  through neon and hydrogen},}\ }\href@noop {} {\bibfield  {journal} {\bibinfo
  {journal} {J. Chem. Phys.}\ }\textbf {\bibinfo {volume} {90}},\ \bibinfo
  {pages} {1007--1023} (\bibinfo {year} {1989})}\BibitemShut {NoStop}%
\bibitem [{\citenamefont {Parravicini}\ and\ \citenamefont
  {Jagau}(2021)}]{parravicini2021embedded}%
  \BibitemOpen
  \bibfield  {author} {\bibinfo {author} {\bibfnamefont {V.}~\bibnamefont
  {Parravicini}}\ and\ \bibinfo {author} {\bibfnamefont {T.-C.}\ \bibnamefont
  {Jagau}},\ }\bibfield  {title} {\enquote {\bibinfo {title} {Embedded
  equation-of-motion coupled-cluster theory for electronic excitation,
  ionisation, electron attachment, and electronic resonances},}\ }\href@noop {}
  {\bibfield  {journal} {\bibinfo  {journal} {Mol. Phys.}\ }\textbf {\bibinfo
  {volume} {119}},\ \bibinfo {pages} {e1943029} (\bibinfo {year}
  {2021})}\BibitemShut {NoStop}%
\bibitem [{\citenamefont {Perdew}, \citenamefont {Burke},\ and\ \citenamefont
  {Ernzerhof}(1996)}]{perdew1996generalized}%
  \BibitemOpen
  \bibfield  {author} {\bibinfo {author} {\bibfnamefont {J.~P.}\ \bibnamefont
  {Perdew}}, \bibinfo {author} {\bibfnamefont {K.}~\bibnamefont {Burke}}, \
  and\ \bibinfo {author} {\bibfnamefont {M.}~\bibnamefont {Ernzerhof}},\
  }\bibfield  {title} {\enquote {\bibinfo {title} {Generalized gradient
  approximation made simple},}\ }\href@noop {} {\bibfield  {journal} {\bibinfo
  {journal} {Phys. Rev. Lett.}\ }\textbf {\bibinfo {volume} {77}},\ \bibinfo
  {pages} {3865} (\bibinfo {year} {1996})}\BibitemShut {NoStop}%
\bibitem [{\citenamefont {Adamo}\ and\ \citenamefont
  {Barone}(1999)}]{adamo1999toward}%
  \BibitemOpen
  \bibfield  {author} {\bibinfo {author} {\bibfnamefont {C.}~\bibnamefont
  {Adamo}}\ and\ \bibinfo {author} {\bibfnamefont {V.}~\bibnamefont {Barone}},\
  }\bibfield  {title} {\enquote {\bibinfo {title} {Toward reliable density
  functional methods without adjustable parameters: The pbe0 model},}\
  }\href@noop {} {\bibfield  {journal} {\bibinfo  {journal} {J. Chem. Phys.}\
  }\textbf {\bibinfo {volume} {110}},\ \bibinfo {pages} {6158--6170} (\bibinfo
  {year} {1999})}\BibitemShut {NoStop}%
\bibitem [{\citenamefont {Lee}, \citenamefont {Yang},\ and\ \citenamefont
  {Parr}(1988)}]{Parr88_785}%
  \BibitemOpen
  \bibfield  {author} {\bibinfo {author} {\bibfnamefont {C.}~\bibnamefont
  {Lee}}, \bibinfo {author} {\bibfnamefont {W.}~\bibnamefont {Yang}}, \ and\
  \bibinfo {author} {\bibfnamefont {R.~G.}\ \bibnamefont {Parr}},\ }\bibfield
  {title} {\enquote {\bibinfo {title} {Development of the colle-salvetti
  correlation-energy formula into a functional of the electron density},}\
  }\href@noop {} {\bibfield  {journal} {\bibinfo  {journal} {Phys. Rev. B}\
  }\textbf {\bibinfo {volume} {37}},\ \bibinfo {pages} {785--789} (\bibinfo
  {year} {1988})}\BibitemShut {NoStop}%
\bibitem [{\citenamefont {Becke}(1988)}]{Becke88_3098}%
  \BibitemOpen
  \bibfield  {author} {\bibinfo {author} {\bibfnamefont {A.~D.}\ \bibnamefont
  {Becke}},\ }\bibfield  {title} {\enquote {\bibinfo {title}
  {Density-functional exchange-energy approximation with correct asymptotic
  behavior},}\ }\href@noop {} {\bibfield  {journal} {\bibinfo  {journal} {Phys.
  Rev. A}\ }\textbf {\bibinfo {volume} {38}},\ \bibinfo {pages} {3098--3100}
  (\bibinfo {year} {1988})}\BibitemShut {NoStop}%
\bibitem [{\citenamefont {Claudino}\ and\ \citenamefont
  {Mayhall}(2019{\natexlab{b}})}]{claudino2019automatic}%
  \BibitemOpen
  \bibfield  {author} {\bibinfo {author} {\bibfnamefont {D.}~\bibnamefont
  {Claudino}}\ and\ \bibinfo {author} {\bibfnamefont {N.~J.}\ \bibnamefont
  {Mayhall}},\ }\bibfield  {title} {\enquote {\bibinfo {title} {Automatic
  partition of orbital spaces based on singular value decomposition in the
  context of embedding theories},}\ }\href@noop {} {\bibfield  {journal}
  {\bibinfo  {journal} {J. Chem. Theory Comput.}\ }\textbf {\bibinfo {volume}
  {15}},\ \bibinfo {pages} {1053--1064} (\bibinfo {year}
  {2019}{\natexlab{b}})}\BibitemShut {NoStop}%
\bibitem [{\citenamefont {Eizner}\ \emph {et~al.}(2019)\citenamefont {Eizner},
  \citenamefont {Martínez-Martínez}, \citenamefont {Yuen-Zhou},\ and\
  \citenamefont {Kéna-Cohen}}]{Eizner2019}%
  \BibitemOpen
  \bibfield  {author} {\bibinfo {author} {\bibfnamefont {E.}~\bibnamefont
  {Eizner}}, \bibinfo {author} {\bibfnamefont {L.~A.}\ \bibnamefont
  {Martínez-Martínez}}, \bibinfo {author} {\bibfnamefont {J.}~\bibnamefont
  {Yuen-Zhou}}, \ and\ \bibinfo {author} {\bibfnamefont {S.}~\bibnamefont
  {Kéna-Cohen}},\ }\bibfield  {title} {\enquote {\bibinfo {title} {Inverting
  singlet and triplet excited states using strong light-matter coupling},}\
  }\href@noop {} {\bibfield  {journal} {\bibinfo  {journal} {Sci. Adv.}\
  }\textbf {\bibinfo {volume} {5}},\ \bibinfo {pages} {eaax4482} (\bibinfo
  {year} {2019})}\BibitemShut {NoStop}%
\bibitem [{\citenamefont {Wu}, \citenamefont {Yan},\ and\ \citenamefont
  {Lalanne}(2021)}]{wu2021}%
  \BibitemOpen
  \bibfield  {author} {\bibinfo {author} {\bibfnamefont {T.}~\bibnamefont
  {Wu}}, \bibinfo {author} {\bibfnamefont {W.}~\bibnamefont {Yan}}, \ and\
  \bibinfo {author} {\bibfnamefont {P.}~\bibnamefont {Lalanne}},\ }\bibfield
  {title} {\enquote {\bibinfo {title} {Bright plasmons with cubic nanometer
  mode volumes through mode hybridization},}\ }\href@noop {} {\bibfield
  {journal} {\bibinfo  {journal} {ACS Photonics}\ }\textbf {\bibinfo {volume}
  {8}},\ \bibinfo {pages} {307--314} (\bibinfo {year} {2021})}\BibitemShut
  {NoStop}%
\end{thebibliography}%

\end{document}


\author{Fabijan Pavo\v{s}evi\'{c}}
\email{fpavosevic@gmail.com}
\affiliation{Center for Computational Quantum Physics, Flatiron Institute, 162 5th Ave., New York, 10010  NY,  USA}

\author{Angel Rubio}
\email{angel.rubio@mpsd.mpg.de}
\affiliation{Nano-Bio Spectroscopy Group and European Theoretical Spectroscopy Facility (ETSF), Universidad del Pa\'is Vasco (UPV/EHU), Av. Tolosa 72, 20018 San Sebastian, Spain}
\affiliation{Max Planck Institute for the Structure and Dynamics of Matter and
Center for Free-Electron Laser Science \& Department of Physics,
Luruper Chaussee 149, 22761 Hamburg, Germany}
\affiliation{The Hamburg Center for Ultrafast Imaging, Luruper Chaussee 149, 22761 Hamburg, Germany}
\affiliation{Center for Computational Quantum Physics, Flatiron Institute, 162 5th Ave., New York, 10010  NY,  USA}

\title[]
%
{Supplemental Information: Wavefunction-in-DFT Embedding for Molecular Polaritons}






\maketitle

\newpage
\section{Reaction Energy Barrier and Reaction Energy for Methyl Transfer in Pyridine with Methyl Bromide Inside an Optical Cavity with Cavity Parameters $\lambda=0.1$ a.u. and $\omega=3$ eV}

\begin{table}[h!]
\caption{Reaction Energy Barrier (TS)\textsuperscript{a} and Reaction Energy ($\Delta E$)\textsuperscript{b} (in kcal/mol) for Methyl Transfer in Pyridine with Methyl Bromide inside an Optical Cavity.}
\centering
\begin{tabular}{c | c c | c c | c c | c c}
\hline
 method & \multicolumn{2}{c|}{outside cavity} & \multicolumn{2}{c|}{$x$ direction} & \multicolumn{2}{c|}{$y$ direction} & \multicolumn{2}{c}{$z$ direction} \\
 \hline\hline
 &      TS    &     $\Delta E$     & TS    &     $\Delta E$     &     TS      &     $\Delta E$     &     TS      &      $\Delta E$    \\
  \hline
QED-HF           & 13.63 & -18.65 & 14.02 & -19.49 & 13.41 & -20.53 & 18.63 & -17.64\\
QED-PBE          & 7.37  & -16.06 & 9.00  & -15.68 & 7.19  & -17.81 & 12.75 & -23.91\\
QED-PBE0         & 7.79  & -19.07 & 8.65  & -19.58 & 7.41  & -21.08 & 15.86 & -21.45\\
QED-B3LYP        & 6.74  & -19.54 & 7.80  & -19.84 & 6.39  & -21.48 & 14.24 & -22.64\\
QED-CCSD         & 10.46 & -21.17 & 9.96  & -22.72 & 10.08 & -22.55 & 13.91 & -23.58\\
QED-CCSD-in-HF   & 10.25 & -23.95 & 10.03  & -25.29 & 9.83  & -25.65 & 13.94 & -26.04\\
QED-CCSD-in-PBE  & 11.56 & -18.86 & 10.91 & -20.75 & 11.11 & -20.58 & 16.23 & -20.05\\
QED-CCSD-in-PBE0 & 11.58 & -19.80 & 11.16 & -21.43 & 11.17 & -21.46 & 15.63 & -21.68\\
QED-CCSD-in-B3LYP& 11.90 & -19.27 & 11.43 & -20.93 & 11.47 & -20.97 & 15.99 & -21.01\\
QED-CCSD-in-PBE0$^{\text{c}}$ & 10.22 & -21.56 & - & - & - & - & - & -\\
\hline
\end{tabular}\\
\textsuperscript{a}\small Relative energies are calculated as the difference between the reaction barrier obtained with the QED method and the corresponding conventional electronic structure method. \\

\textsuperscript{b}\small Relative energies are calculated as the difference between the reaction energy (i.e., the difference between the energies of the product and reactant) obtained with the QED method and the corresponding conventional electronic structure method.\\

\textsuperscript{c}\small Embedding domain that include additional two adjacent CH groups.\\
\label{table:reaction_profile_SN2}
\end{table}

\newpage

\section{Reaction Energy Barrier and Reaction Energy for Proton Transfer in Aminopropenal Inside an Optical Cavity with Cavity Parameters $\lambda=0.1$ a.u. and $\omega=3$ eV}

\begin{table}[h!]
\caption{Reaction Energy Barrier (TS)\textsuperscript{a} and Reaction Energy ($\Delta E$)\textsuperscript{b} (in kcal/mol) for Methyl Transfer in Pyridine with Methyl Bromide inside an Optical Cavity.}
\centering
\begin{tabular}{c | c c | c c | c c | c c}
\hline
 method & \multicolumn{2}{c|}{outside cavity} & \multicolumn{2}{c|}{$x$ direction} & \multicolumn{2}{c|}{$y$ direction} & \multicolumn{2}{c}{$z$ direction} \\
 \hline\hline
 &      TS    &     $\Delta E$     & TS    &     $\Delta E$     &     TS      &     $\Delta E$     &     TS      &      $\Delta E$    \\
  \hline
QED-HF           & 16.65 & 8.63 & 18.65 & 9.57 & 17.24 & 9.00 & 16.64 & 8.69\\
QED-PBE          & 3.93  & 6.26 & 6.14  & 7.07 & 4.08  & 6.08 & 3.83  & 6.26\\
QED-PBE0         & 6.09  & 6.62 & 8.36  & 7.45 & 6.45  & 6.59 & 6.02  & 6.63\\
QED-B3LYP        & 7.32  & 6.80 & 9.58  & 7.62 & 7.64  & 6.73 & 7.25  & 6.81\\
QED-CCSD         & 9.65  & 4.54 & 10.72 & 4.67 & 9.71  & 4.32 & 9.32  & 4.34\\
QED-CCSD-in-HF   & 10.43 & 5.49 & 11.75 & 5.59 & 10.43 & 5.07 & 10.16 & 5.16\\
QED-CCSD-in-PBE  & 11.06 & 6.38 & 12.78 & 6.07 & 11.04 & 6.08 & 10.77 & 6.18\\
QED-CCSD-in-PBE0 & 10.74 & 5.95 & 12.29 & 5.80 & 10.71 & 5.63 & 10.46 & 5.74\\
QED-CCSD-in-B3LYP& 11.29 & 5.87 & 12.93 & 5.66 & 11.26 & 5.53 & 11.01 & 5.65\\
\hline
\end{tabular}\\
\textsuperscript{a}\small Relative energies are calculated as the difference between the reaction barrier obtained with the QED method and the corresponding conventional electronic structure method. \\

\textsuperscript{b}\small Relative energies are calculated as the difference between the reaction energy (i.e., the difference between the energies of the product and reactant) obtained with the QED method and the corresponding conventional electronic structure method.\\

\label{table:reaction_profile_aminopropenal}
\end{table}

\newpage
\section{Cartesian Coordinates of the Optimized Geometries}

Optimized with MP2/6-31G(d) method:\\ 
Pyridine with Methyl Bromide Reactant (Number of Imaginary Frequencies=0)
\begin{verbatim}
 C                  1.13759681   -0.05477211   -2.19306978
 H                  0.24449380    0.44574240   -2.55396116
 H                  1.13314176   -1.09244459   -2.51982225
 H                  2.01600197    0.44796781   -2.58556594
 Br                 1.08790491   -0.22514853   -5.15588397
 N                  1.16435647   -0.00324395   -0.68355764
 C                  0.00000000    0.00000000    0.00000000
 C                  0.00000000    0.00000000    1.38825264
 C                  1.21359614    0.00000000    2.07545441
 C                  2.40191143   -0.00274236    1.34538743
 C                  2.35238160   -0.00267165   -0.04196715
 H                  1.23297434    0.00214095    3.16119421
 H                 -0.90284719   -0.00066197   -0.60145425
 H                 -0.94822464    0.00436088    1.91541110
 H                  3.36834811   -0.00056812    1.83838151
 H                  3.23316883   -0.00536202   -0.67529613
\end{verbatim}

Optimized with MP2/6-31G(d) method:\\
Pyridine with Methyl Bromide Transition State (Number of Imaginary Frequencies=1)
\begin{verbatim}
 C                  1.10026781   -0.05950199   -2.53988699
 H                  1.09727598   -1.13283765   -2.63133480
 H                  2.02255209    0.47277974   -2.70410122
 H                  0.17128876    0.47383242   -2.65586161
 Br                 1.03462545   -0.13090326   -5.05746718
 N                  1.14831147   -0.00003940   -0.69694831
 C                  0.00000000    0.00000000    0.00000000
 C                  0.00000000    0.00000000    1.39173587
 C                  1.22030589    0.00000000    2.06832316
 C                  2.40375234   -0.00091860    1.32915551
 C                  2.33133285   -0.00089199   -0.06069652
 H                  1.24857876    0.00072777    3.15428244
 H                 -0.91605046   -0.00071297   -0.58576567
 H                 -0.94245073    0.00142231    1.93058869
 H                  3.37296670   -0.00022807    1.81823783
 H                  3.21566118   -0.00228926   -0.69333534
\end{verbatim}

Optimized with MP2/6-31G(d) method:\\
Pyridine with Methyl Bromide Product (Number of Imaginary Frequencies=0)
\begin{verbatim}
 C                  1.02368319    0.05729769   -3.88264194
 H                  0.14367313    0.58843316   -3.53298410
 H                  1.02352551   -0.96949641   -3.53110392
 H                  1.93068036    0.57179520   -3.58034422
 Br                 0.97146917    0.02891100   -5.84054942
 N                  1.12024672    0.00002885   -0.74573812
 C                  0.00000000    0.00000000    0.00000000
 C                  0.00000000    0.00000000    1.39525820
 C                  1.22336316    0.00000000    2.06448953
 C                  2.39439815   -0.00016595    1.30740911
 C                  2.29212447   -0.00016307   -0.08409104
 H                  1.26322069   -0.00005113    3.15080347
 H                 -0.93752389   -0.00055095   -0.55341536
 H                 -0.93952305   -0.00007929    1.94067080
 H                  3.37136483   -0.00038187    1.78250172
 H                  3.18659424   -0.00087954   -0.70471099
 \end{verbatim}

Optimized with CCSD/cc-pVDZ method:\\ 
Aminopropenal Reactant (Number of Imaginary Frequencies=0)
\begin{verbatim}
 H        0.00000000    0.00000000    0.00000000
 C        0.00000000    1.09163906    0.00000000
 C        1.26822472    1.80690645    0.00000000
 N       -1.33404678    3.10992888    0.00000000
 O        1.39279087    3.03379031    0.00000000
 H       -0.48254004    3.66582880    0.00000000
 H       -2.24599052    3.54003736    0.00000000
 C       -1.19931821    1.76513995    0.00000000
 H        2.18220599    1.16371496    0.00000000
 H       -2.14055701    1.20128576    0.00000000
\end{verbatim}

Optimized with CCSD/cc-pVDZ method:\\
Aminopropenal Transition State (Number of Imaginary Frequencies=1)
\begin{verbatim}
 H        0.00000000    0.00000000    0.00000000
 C        0.00000000    1.08943910    0.00000000
 C        1.19187861    1.82073020    0.00000000
 N       -1.15503519    3.15085322    0.00000000
 O        1.23956085    3.10723950    0.00000000
 H        0.06714025    3.42106531    0.00000000
 H       -2.03971004    3.65182794    0.00000000
 C       -1.21230682    1.83679656    0.00000000
 H        2.16133262    1.29142618    0.00000000
 H       -2.18458586    1.32025952    0.00000000
\end{verbatim}

Optimized with CCSD/cc-pVDZ method:\\
Aminopropenal Product (Number of Imaginary Frequencies=0)
\begin{verbatim}
 H        0.00000000    0.00000000    0.00000000
 C        0.00000000    1.09138840    0.00000000
 C        1.19000326    1.75971354    0.00000000
 N       -1.29869978    3.11237163    0.00000000
 O        1.33369004    3.08569523    0.00000000
 H        0.41141507    3.45367072    0.00000000
 H       -2.26092059    3.46346742    0.00000000
 C       -1.27014745    1.81619151    0.00000000
 H        2.14625767    1.22260226    0.00000000
 H       -2.19412100    1.21051211    0.00000000
 \end{verbatim}

\clearpage